\documentclass[jgrga,draft]{agujournal}
\journalname{JGR Planets}

\usepackage{graphicx}
\usepackage{amssymb}

\setkeys{Gin}{draft=false}

\begin{document}


\title{Geophysical tests for habitability in ice-covered ocean worlds}




\authors{Steven D. Vance\affil{1}, 
Mark P. Panning\affil{1}\thanks{now at Jet Propulsion Laboratory, California Institute of Technology, Pasadena, California, USA}, 
Simon St\"ahler \affil{2}\thanks{now at Institute of Geophysics, ETH Z\"{u}rich, Z\"{u}rich, Switzerland}
Fabio Cammarano\affil{3}, 
Bruce G. Bills\affil{1}, 
Sharon Kedar\affil{1}, 
Christophe Sotin\affil{1}, 
William T. Pike\affil{4}, 
Ralph Lorenz\affil{5}, 
Victor Tsai\affil{6}, 
Hsin-Hua Huang\affil{6,7}, 
Jennifer M. Jackson\affil{6}, 
Bruce Banerdt \affil{1}}

\affiliation{1}{Jet Propulsion Laboratory, California Institute of Technology, Pasadena, California, USA}
\affiliation{2}{Dept. of Earth and Environmental Sciences, Ludwig-Maximillians-Universit\"{a}t, Munich, Germany}
\affiliation{3}{Dipartimento di Scienze Geologiche, Universita Roma Tre, Largo S. L. Murialdo 1, 00146 Rome, Italy}
\affiliation{4}{Department of Electrical and Electronic Engineering, Imperial College, London, United Kingdom}
\affiliation{5}{Johns Hopkins University Applied Physics Laboratory, Laurel, USA}
\affiliation{6}{Seismological Laboratory, California Institute of Technology, Pasadena, California, USA}
\affiliation{7}{Institute of Earth Sciences, Academia Sinica, Taiwan}




\begin{abstract}

Geophysical measurements can reveal the structure of icy ocean worlds and cycling of volatiles. The associated density, temperature, sound speed, and electrical conductivity of such worlds thus characterizes their habitability.  To explore the variability and correlation of these parameters, and to provide tools for planning and data analyses,  we develop 1-D calculations of internal structure, which use available constraints on the thermodynamics of aqueous MgSO$_4$, NaCl (as seawater), and NH$_3$, water ices, and silicate content. Limits in available thermodynamic data narrow the parameter space that can be explored: insufficient coverage in pressure, temperature, and composition for end-member salinities of MgSO$_4$ and NaCl, and for relevant water ices; and a dearth of suitable data for aqueous mixtures of Na-Mg-Cl-SO$_4$-NH$_3$. For Europa, ocean compositions that are oxidized and dominated by MgSO$_4$, vs reduced (NaCl), illustrate these gaps, but also show the potential for diagnostic and measurable combinations of geophysical parameters. The low-density rocky core of Enceladus may comprise hydrated minerals, or anydrous minerals with high porosity comparable to Earth's upper mantle. Titan's ocean must be dense, but not necessarily saline, as previously noted, and may have little or no high-pressure ice at its base.  Ganymede's silicious interior is deepest among all known ocean worlds, and may contain multiple phases of high-pressure ice, which will become buoyant if the ocean is sufficiently salty. Callisto's likely near-eutectic ocean  cannot be adequately modeled using available data.  Callisto may also lack high-pressure ices, but this cannot be confirmed due to uncertainty in its moment of inertia.

\correspondingauthor{S. D. Vance}{svance@jpl.nasa.gov}
\begin{keypoints}
\item Provides tools for planning and data analysis needed for geophysical investigations of icy ocean worlds
\item Constrains interior structures and explores the influence of composition using self-consistent thermodynamics of ices, oceans, and silicates
\item  Demonstrates the potential for obtaining a combination of geophysical signatures that are diagnostic of conditions that might support life
\end{keypoints}

\end{abstract}

\section{Introduction}


Geophysical measurements can reveal the interior properties of icy ocean worlds. Such measurements  can point to the presence and temporal variability of fluids and gases, thus identifying potential habitable niches for life. The thickness of ices and depth of the ocean determine the conditions under which life might exist.  Similarly, the global fluxes of chemical energy must be understood in terms of a world's interior structure and evolution, which can only be illuminated by geophysical means. Earth's biogeochemical cycles rely on the mantle for a continuous supply of reduced materials \citep[e.g.,][]{hayes2006carbon} balanced by a continuous supply of oxidants \citep[e.g.,][]{catling2005how}. 

Consider the example of Jupiter's moon, Europa. The small value of Europa's gravitational moment of inertia inferred from  \textit{Galileo} flybys \citep[Table~\ref{table:planetProps};][]{anderson1998europas,schubert2004interior} suggests a differentiated body consistent with strong tidal heating expected to have occurred throughout its history \citep{obrien2002meltthrough,sotin2009tides}. It is commonly assumed that Europa's ocean started out hot and reducing, and after some time ($\sim$2~Gyr) may have  oxidized due the flux of radiolytic materials produced at the surface  \citep{pasek2012acidification}. The flux today of reduced hydrogen and other materials from high-temperature hydrothermal activity and water-rock alteration has been estimated at  $>10^{9}$~moles~yr$^{-1}$, with potentially an order of magnitude greater oxidizing flux from materials produced at the surface  \citep{hand2007energy,vance2016geophysical}. 

The example of Europa, analogous to its smaller sibling Enceladus, illustrates how the accumulated effects of an ocean world's redox evolution can be measured by geophysical means. Whether Europa had its own paleozoic (or azoic) "great oxidation event" \citep{lyons2014rise} depends on its initial composition and geodynamic evolution.  The ocean's pH and associated salinity are the integrated results of a world's geochemical evolution; a low pH ocean would be dominated by sulfate anions, whereas a neutral or basic ocean would be dominated by chlorides \citep{zolotov2008oceanic,zolotov2009chemical}. 

Ocean composition measurements cannot be based on surface imaging or atmospheric sampling alone, due to unknown fractionation within the ice. Chlorides inferred from surface infrared reflectance spectra of Europa's surface \citep{brown2013salts,fischer2015spatially,ligier2016vlt} are not a conclusive proxy for seawater salinity, because chlorides fractionate into freezing sea ice relative to sulfates \citep{gjessing1997chemical,maus2011ion}. Direct measurement of an ocean's salinity and characterization of the ice can only be achieved in the near term by geophysical means. 

The habitability of larger icy ocean worlds may also be probed with geophysical measurements. In Ganymede, Callisto, and Titan, high-pressure ices have been regarded as limiting water-rock interactions that regulate redox and ocean composition, and which are thereby critical for supporting life. However, convective transport within high-pressure ices should proceed near the solidus temperature, leading to substantial melt at the ice-rock interface, and within the upper part of the ice \citep{choblet2017heat}.  Briny fluids under pressure can have densities exceeding those of high pressure ices, and may occupy the interface between rock and ice \citep{hogenboom1995magnesium,vance2014ganymede}.  Dissolved ions can incorporate into high-pressure ice VI (and possibly phases II, III, and V) to a much greater extent than in ice~Ih, potentially decreasing the density of the ices and enhancing near-solidus convection \citep{journaux2017salt}.   

Seismic investigations could characterize habitability of ocean worlds \citep{vance2016seismic} in a manner that is highly complementary to other measurements on missions exploring the habitability of ocean worlds, particularly those of NASA's planned Europa Mission \citep{pappalardo2016science} and ESA's JUICE mission \citep{grasset2013JUICE}: Radar investigations will examine the density, temperature, and composition of the ices covering Europa, Ganymede, and Callisto. The RIME and REASON instruments will reveal the upper km of the ice with a range resolution as small as 15 meters, and will sound to as much as 45 km in cold ice with a range resolution as small as 30~m \citep{grima2015radar}. Whereas radio waves are sensitive mainly to density, electrical conductivity, and crystal orientation fabric (COF), seismic waves are influenced mainly by density, COF, and temperature \citep{diez2013effects}. Prior radar information thus provides a powerful constraint for extracting more detailed information from the deeper interior. Similar combined investigations of Antarctic glaciers have revealed sub-ice lake and low-velocity layers at the water-rock interface consistent with a sedimentary layer \citep{bell1998influence}. Electromagnetic measurements (ICEMAG and PIMS) on the planned Europa Clipper mission \citep{pappalardo2016science} will use Europa's induced response to Jupiter to constrain the ocean's depth and electrical conductivity. Seismology could directly measure the ice and ocean thicknesses, and could leverage constraints on electrical conductivity to narrow the range of plausible models for ocean salinity. Seismology provides comparable or better vertical resolution, and a measure of the magnitude and nature of ongoing activity. Seismic measurements might also probe below the ice-ocean interface to reveal the deeper interior,  with the potential to constrain the current thermal state of the rocky interior \citep{cammarano2006longperiod}. 
This combination of seismology with other methods may be deployed at other ocean worlds to reveal the nature of a potential habitat in the ice, ocean, seafloor, and below.

Forward models of the interiors of icy ocean worlds are needed in order to plan future mission designs and measurements of their interior dynamics and habitability. Here, we develop global one-dimensional models to assess how combined seismic, gravity, and magnetic measurements may be used to investigate the habitability of ocean worlds.  We use the models to consider how the ocean's composition might influence the configuration and thickness of ice layers. Another important function of this work is to test the limits of available data describing material properties. 

In Section~\ref{section:methods}, we describe the workings of the model and limitations of available data.  In Section~\ref{section:results}, we describe applications to confirmed ocean worlds, Ganymede, Europa, Enceladus,  Titan, and Callisto. In Sections~\ref{section:discussion}~and~\ref{section:conclusions} we discuss possible tests for habitability, and  thermodynamic measurements needed for further modeling.


\section{Materials and Methods}\label{section:methods}
We assess interior density, elastic and anelastic structure,and associated seismic sources and signatures, building on prior work that focused only on Europa \citep{cammarano2006longperiod}. Added to this is an investigation of the electrical conductivity of fluids based on available measurements.

The interior structure model propagates density and temperature profiles downward from the ice I layer from boundary conditions of the surface and ice-ocean boundary temperature, as per \citet[][; Appendix \ref{appendixMD}]{vance2014ganymede}. Significant improvements have been made for speed, modularity to enable future work, and added geodynamics and geochemistry for the ice I layer and deeper interior. Depths of transition between compositional layers \{ice, ocean, (ice), rock, (iron)\} are computed from bulk density and gravitational moment of inertia ($C/MR^2$; Table~\ref{table:planetProps}).  Solid state convection in the ice~I layer has been added, and is included by iteration after the initial model run. Gravitational acceleration increases with depth in the relatively low density ice and liquid layers approaching the solid core \citep[by up to 25\% in the case of Ganymede;][]{vance2014ganymede}, so depth-dependent gravity is computed after the initial computation of the pressure-dependent adiabatic gradient. Slowly rotating body Callisto may not be in hydrostatic equilibrium \citep{gao2013nonhydrostatic} and so the  moment of inertia inferred from simultaneous measurements of \textit{J$_2$} and \textit{C$_{22}$}  may be overestimated by as much as 10\%. We consider this effect in the Section \ref{section:results}.

\begin{table*}[h]
\caption{Properties of ocean worlds}
\begin{tabular}{|c|c|c|c|}
\hline
 & Radius (km) & Density (kg m$^{-3}$) &  Moment of Inertia\\
\hline
Europa $^a$& 	1565.0$\pm8.0$			& 2989$\pm$46		&	0.346$\pm0.005$\\
Ganymede$^a$ & 	2631$\pm1.7$	& 1942.0$\pm4.8$	&	0.3115$\pm0.0028$\\
Callisto$^a$ & 	2410.3$\pm1.5$	& 1834.4$\pm3.4$	&	0.3549$\pm0.0042$\\
Enceladus$^{b,c}$   & 252.1$\pm$0.2			& 1609$\pm$5				&	0.335\\
Titan$^c$& 	2574.73$\pm0.09$& 1879.8$\pm0.004$				&   0.3438$\pm0.0005$\\
\hline
\multicolumn{4}{l}{$^{a}$\citet{schubert2004interior}, $^{b}$\citet{thomas2010sizes,iess2014gravity}, $^{c}$\citet{jacobson2006gfs,iess2012tides}}
\end{tabular}
\label{table:planetProps}
\end{table*}
 
\subsection{Thermodynamics and Geodynamics}
The model's current modules are described in the following subsections. Thermodynamic modules provide self-consistent density, thermal expansivity, specific heat capacity, and sound speed constrained by available data, using published methodologies to easily and rapidly retrieve values. Adding new equations of state for alternate ocean compositions or updated material properties is straightforward.

\subsubsection{Ices}
\begin{figure*}[h]
\noindent\includegraphics[width=45pc]{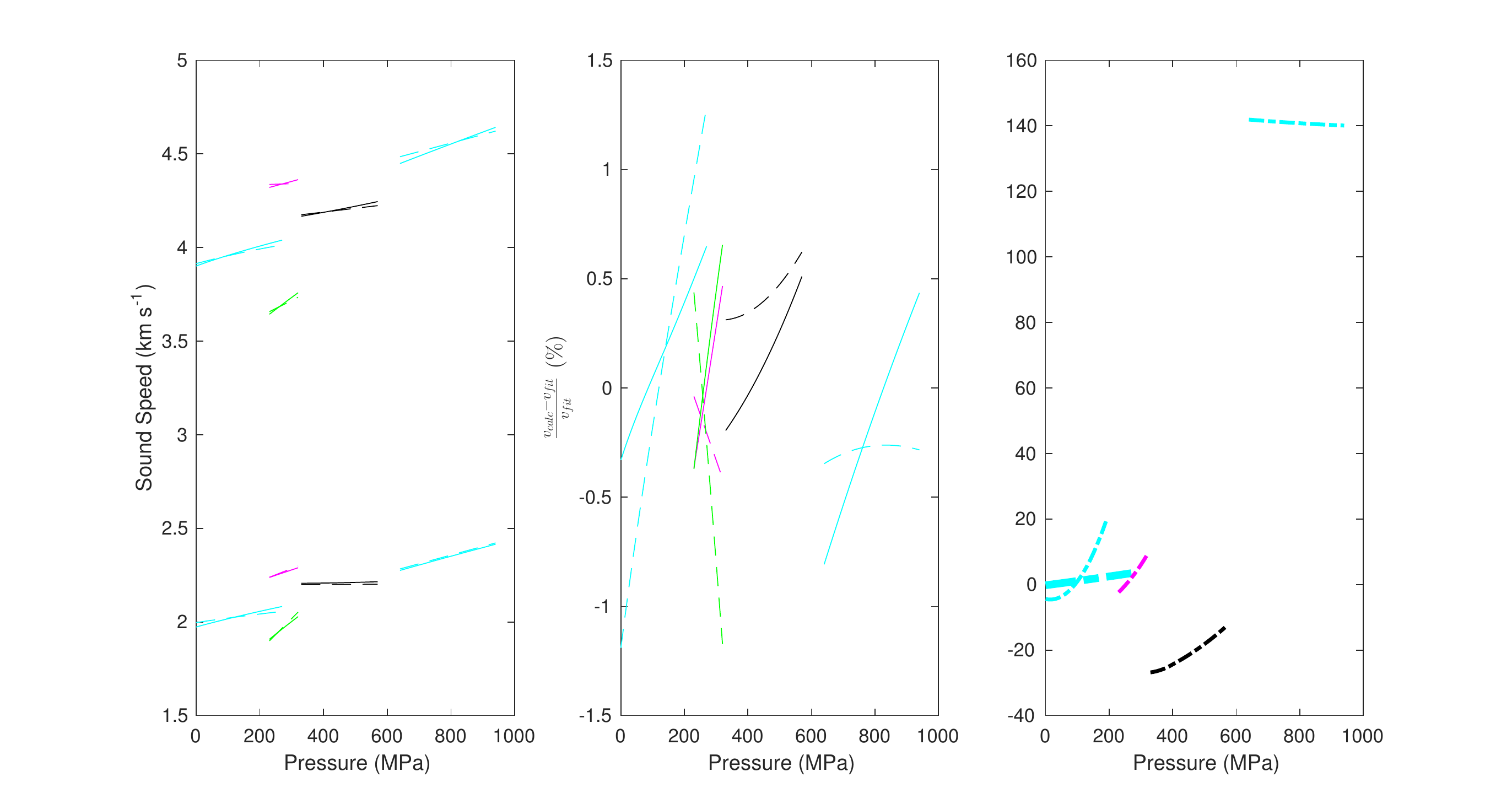}
\caption{Evaluation of sound speeds near 240~K.  Sound speeds (left) from \citep{gagnon1988pressure} and \citep{gagnon1990acoustic} (--) compared with the fit used in this work.  Middle: deviations (\%) between published and fit values ($V_P$ --, $V_S$ -- --). Right: Misfit (\%) between bulk sound speeds computed from published  values and compiled thermodynamics \citep{choukroun2010thermodynamic}.}
\label{figure:vel35}
\end{figure*}

\begin{figure*}[h]
\noindent\includegraphics[width=45pc]{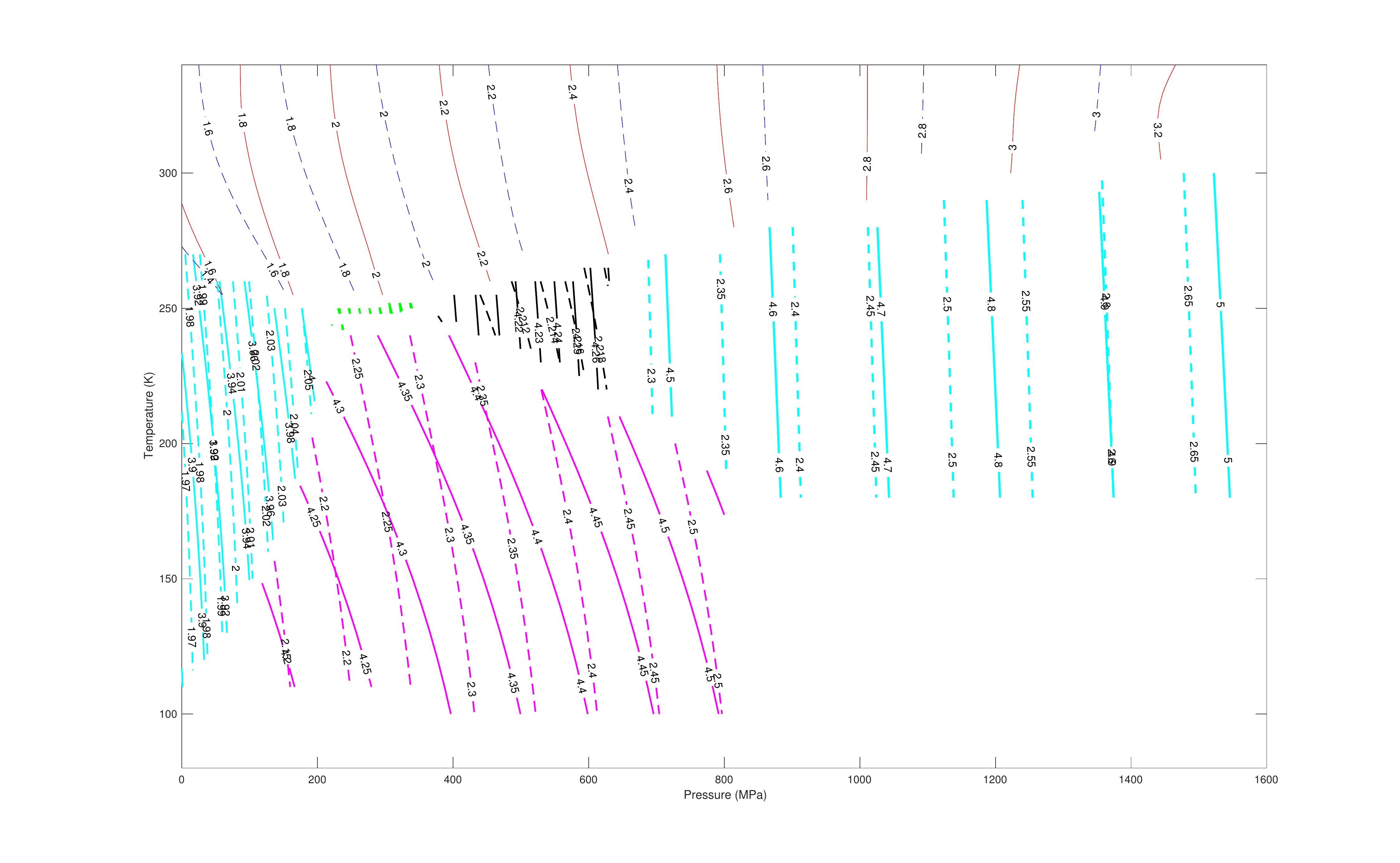}
\caption{Contours of sound speed versus pressure (MPa) and temperature (K). In the ice stability field ($T\sim<260$~K): $v_P$ (--) and $v_S$ (- -). In the fluid stability range: $v_P$ in 10~wt\% MgSO$_4$ (--) and pure water (- -). Fluid-liquid phase boundaries are for the 10~wt\% case.}
\label{figure:soundContours}
\end{figure*}

The thermodynamic properties of water ice phases Ih, II, III, V, and VI  are taken primarily from \citet{choukroun2010thermodynamic}, which fit available measurements of heat capacity, enthalpy, and specific volume over the broad range of pressure and temperature occurring in icy worlds. Corresponding bulk sound speeds ($v_{C}$) ($(\partial \rho/\partial P)_T=v_{C}^{-2}+\alpha^2 T/C_P$) have very large pressure derivatives and misfit available measurements by up to 100\% (right-hand pane in Figure~\ref{figure:vel35}).  To provide sound speeds consistent with available measurements and mostly consistent with thermodynamic data, compressional and transverse sound speeds in these phases are matched to sound speeds from bulk adiabatic Brillouin measurements in polycrystalline ice at $-35.5^{o}C$ to 1~GPa \citep{gagnon1988pressure,gagnon1990acoustic} to within their assigned errors. Adiabatic shear and bulk moduli ( $\mu$, $K_S$) and their pressure derivatives were adjusted from \citet{shaw1986elastic} to fit compiled sound speeds from \citet{gagnon1990acoustic} to better than 1.5\%. Sound speeds (Figure~\ref{figure:soundContours}) are computed as
\begin{eqnarray}
V_S=\bigg[\frac{\mu}{\rho}\bigg]^{1/2}\\
V_P=\bigg[\frac{K_S}{\rho}+\frac{4}{3}V_S^2\bigg]^{1/2}
\end{eqnarray} 
This strategy does not reproduce the negative slope of sound speed with temperature of $\approx-2$~m~s$^{-1}$~K$^{-1}$ \citep{vogt2008speed}. Further measurements of sound velocities in ices would aid both seismic data analysis and geodynamics.

\begin{table}[h]
\caption{Bulk and Shear Moduli for Ices}
\centering
\begin{tabular}{|c|c|c|c|c|c|c|}
\hline
Crystalline   & $K_S$$^a$ & $K_S'$$^b$&$K_S''$&$\mu$ & $\mu'$$^b$  &$\mu''$ \\
Phase  		  &  GPa & &GPa$^{-1}$&GPa & &GPa$^{-1}$\\ 
\hline
I$_h$$^c$ & 9.5	&0.33&-0.026&3.3&0.537&-.025\\
II & 		13.89	&1.6&---&5.15&3.5&---\\
III & 		8.9	&3.65&--- &2.7&6.55&---\\
V   & 		11.8	&4.8&---&5.7&0.9&---\\
VI  & 		14.6	&4.1&---&5.0&3.0&---\\
\hline
\multicolumn{7}{l}{
$^a$Adapted from \citet{gagnon1990acoustic},$^b$Fit from \citet{shaw1986elastic},$^c$\citet{gagnon1988pressure}}
\end{tabular}
\label{table:materialProps}
\end{table}

\subsubsection{Fluids}
We examine the influence of ocean salinity in each ocean world by comparing compositions of pure water and 10~wt\% ($\sim$1 molal) MgSO$_4$. Seawater compositions are examined for Europa and Enceladus. Solutions containing 3~wt\% NH$_3$ are considered for Titan.

The thermodynamic module for the fluids (\texttt{fluidEOS}) provides properties for aqueous magnesium sulfate from \citet{vance2013thermodynamicMgSO4}. Properties are retrieved by three-dimensional spline interpolation, with extrapolation above 0.8~GPa and 2.5~mol~kg$^{-1}$. The melting point depression is computed using the Margules parameterization for the activity of water for MgSO$_4$ \citet{vance2014ganymede} and NH$_3$ \citet{choukroun2010thermodynamic}. Contours of sound speed for water and 10~wt\%~MgSO$_4$ are shown in Figure~\ref{figure:soundContours}.The Gibbs Seawater package \citep[GSW;][]{mcdougall2011getting} provides self-consistent thermodynamics for seawater reliably to 100~MPa, with reasonable stability for standard concentration to the 210~MPa limitation of its equation of state for ice Ih. Thermodynamics of aqueous ammonia are from \citep{tillner1998helmholtz} with phase equilibria from .

\begin{figure}[h]
\noindent\includegraphics[width=20pc]{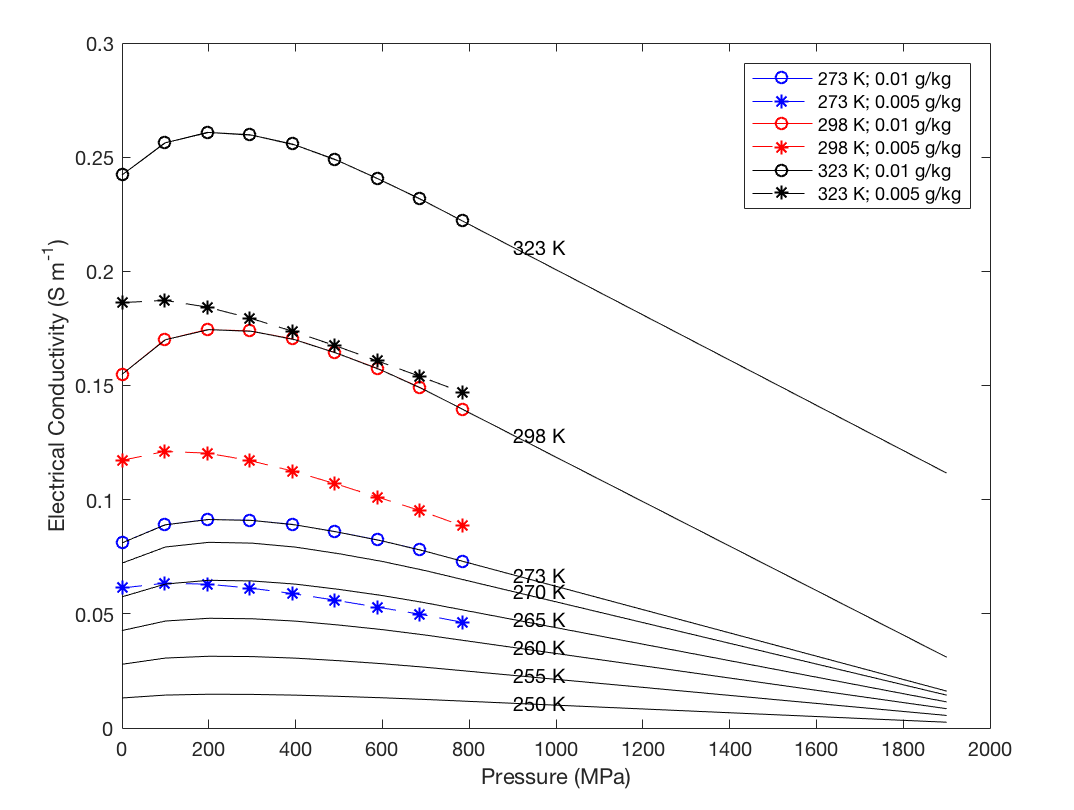}
\caption{Electrical conductivity of 0.005 molal and 0.01 molal MgSO$_4$ from \citet{larionov1984conductivity}. Contours show linear extrapolation above 800~MPa and below 273~K for 1 molal conductivity.}
\label{figure:conductivity}
\end{figure}

The electrical conductivity of 0.01~mol~kg$^{-1}_{H_{2}O}$ (molal) magnesium sulfate (aq) is computed along geotherms based on measurements at 298~K and 323~K, and pressures up to 784.6~MPa \citep{larionov1984conductivity}. Extrapolation to 273~K is applied with a scaling factor of 0.525 as per \citet{hand2007empirical}, with linear interpolation and extrapolation to 250~K and 1.6~GPa. Figure~\ref{figure:conductivity} shows the variation of electrical conductivity with pressure, temperature, and concentration in the main region down to 273~K. Conductivity depends strongly on concentration, but direct measurements at elevated pressure are unavailable.   Measurements taken at standard pressure show a factor of 40 increase for magnesium sulfate up to 1 molal  \citep[][Figure~1]{hand2007empirical}. This linear scaling  with concentration is used in our calculations. GSW includes electrical conductivity.

\subsubsection{Rock}
We model the silicate portions of planetary bodies using available thermodynamic software. Stable mineralogies, densities, and seismic velocities of the bulk rock  are determined from the profiles of present-day temperature ($T$), composition ($C$), and pressure ($P$) for a range of possible compositions using PERPLEX~6.7.3 \citep{connolly2005computation,connolly2009geodynamic,perplex}. The program  solves the Gibbs  energy minimization based on specified thermodynamic databases and thermoelastic properties for end-member mineralogical phases. Average properties of the bulk rock are determined through classical Voigt-Reuss-Hill averaging scheme.  High-pressure ($>$5GPa) polythermal equations of state (Murnaghan, Birch-Murnaghan) were stabilized by computing the temperature derivative of the bulk modulus as a function of the pressure derivative of the bulk modulus and the expansivity at the reference pressure, as per \citet{helffrich2009physical}.

\subsubsection{Bulk Mineralogy}
Following previous work for Europa by \citet{cammarano2006longperiod}, we consider two different dry compositions: an Earth-like pyrolite composition and L-LL~chondritic model (Table~\ref{table:chondrite}). We also consider updated compositions that include sodium, as per \citet{cammarano2011seismic}. The computations use the formalism and thermoelastic properties dataset of \citet{stixrude2011thermodynamics}, which accounts for six main oxides: SiO$_2$, MgO, FeO, CaO, Al$_2$O$_3$, Na$_2$O.  The database was assembled with a focus on using available elastic moduli of Earth's mantle minerals  extending to pressures in the 100~GPa range.
Variations in major oxides, especially in iron content, have the largest effect on density. The range of densities within the plausible range of pressure-temperature conditions covering all icy moons is small compared with Earth's mantle. Densities for both pyrolite and L-LL~chondrite composition exceed 3000 kg~m$^{-3}$. This is also true for larger variations in density arising from changes in oxide composition.

\subsubsection{Hydrous Minerals}
The effects of hydrous compositions are also considered,  using the database of \citet{holland2011improved}, which includes more mineral species while retaining thermodynamic self-consistency. We use the same  compositions described in the previous section, but at partially hydrated or water-saturated conditions. Serpentinization may be important as a source of H$_2$ for life \citep[e.g.,][]{muntener2010serpentine,vance2016geophysical}, but it
may not reduce the density of the rocky siliceous parts of icy worlds because hydrated silicates are not thermodynamically stable at low pressures, and at temperatures higher than 900~K \citep[e.g.,][]{ulmer1995serpentine}.

\subsubsection{Porosity}
Because of the low confining pressures occurring in  smaller icy ocean worlds, porosity becomes important at much greater depths than on Earth. Tidal forcing and thermal fracturing may aid in developing a porous interior, as putatively found at Enceladus \citep{vance2007hydrothermal}.
In general, porosity decreases with increasing lithostatic pressure, which closes pore spaces, cracks and fractures.
\citet{vitovtova2014porosity}, for example, estimated Earth's mantle porosity at a few percent at a depth of 10~km, decreasing to 0.01-0.1\% at 35~km. 
However the trend can be more complex. For example an increase in porosity with depth is recorded for the borehole of the Kola peninsula, the deepest on Earth, at 12 km \citep{kozlovsky1987geothermic}. 

Direct and indirect measurements suggest the porosity of Earth's oceanic and continental crust is large, especially in the uppermost part \citep[ranging from 5 to 10\%, e.g.][]{carlson2014influence}. For example, high $V_P/V_S$ at 25-45 km depth suggest a porosity around 2.7-4.0 \%  \citep{peacock2011high}, associated with the presence of fluid saturated zones. Alternatively, crack anisotropy may partially explain the high $V_P/V_S$ \citep{wang2012high}. Experimental studies find that voids are closed at pressures higher than 0.25 GPa \citep{kern1990laboratory} unless fluids fill the voids. Recent laboratory experiments in a porous natural sample of crustal rock \citep[\i.e.,][]{saito2016laboratory} indicate that porosity closure is active up to $\sim$0.6 GPa. In spite of low remanent porosity, the effect on seismic velocities is a few percent, and thus should be accounted for. The $V_P/V_S$ ratio, in this case, decreases in the pore spaces not filled by fluids.
In crystalline rocks, the intergranular porosity is usually small, but may increase due to water-rock interactions \citep[e.g.,][]{tutolo2016nanoscale}. 

We model the decrease of porosity with pressure using the empirical formula provided by \citet{vitovtova2014porosity}: $log(\phi) = - 0.65 - 0.1d + 0.0019 d^2$, where $\phi$ is the total porosity (assumed to be equal to the effective one) and $d$ the depth in km. We convert depth to pressure using the pressure profile from the preliminary reference earth model  \citep[PREM;][]{dziewonski1981preliminary}.
The reduction in density due to porosity is obtained from $\rho=\phi \rho_f+ (1-\phi) \rho_g$, where the rock density $\rho_g$ is the obtained density from thermodynamic modeling.  The pore spaces are assumed to be filled by water (\textit{i.e.}, $\rho_f=1023$ kg m$^{-3}$). 

Concerning the effects of porosity on seismic velocities, few experiments exist on the effects of low porosity crystalline rocks. \citep[e.g.,][]{todd1972effect,christensen1989pore,darot2000effect,yu2016effects}. In general, the effect of pore pressure on seismic velocities is  inverse to that of the confining pressure; 
 seismic velocities are expected to decrease as pore pressures increase. The expected behavior is complicated by the shape, distribution and connectivity of pores, and the presence of cracks and fractures.
We account for the decrease in $V_P$ as a function of porosity using the relation suggested by \citet{wyllie1958experimental}. The decrease in $V_S$ should be greater; \citet{christensen1989pore} found a decrease of 9\% and 26\% for $V_P$ and $V_S$, respectively, when increasing pore pressure from atmospheric pressure to 85\% in a very porous (3.9\%) lherzolite xenolith at 150 MPa of confining pressure. We scale the decrease in $V_S$ to be 2 times larger in percentage than $V_P$. Although  approximate, these corrections account for the expected qualitative effects of water-filled porous spaces on seismic velocities.


For models specifying an iron core, the starting density of the rock is adjusted to match the average predicted by the model.

\begin{table}[h]
\caption{Chemical compositions (in mol\%) considered for the rocky component}
\centering
\begin{tabular}{|c|c|c|c|c|}
\hline
	&Pyrolite$^a$ & L-LL Chondrite$^b$  & Pyrolite$^b$ & L-LL Chondrite$^b$ \\
\hline
SiO$_2$ &38.71	&  38.73	&	38.66	&	42.78				\\
MgO & 	49.85	&  38.73	&	48.53	&	39.68			\\
FeO & 	6.17	&  14.98	&	5.72	&	13.98			\\
CaO & 	2.94	&  2.12		&	3.50	&	2.13			\\
Al$_2$O$_3$ & 2.22&1.36		&	3.59	&	1.43			\\
Na$_2$O & 0.11	&  1.02		&	 --	    &	--	     	\\
\hline
\multicolumn{5}{l}{$^a$\citet{cammarano2011seismic};$^b$modified from \citet{cammarano2006longperiod}}
\end{tabular}
\label{table:chondrite}
\end{table}
 
\subsubsection{Iron Core}
When including an iron core, we assume the $\gamma$ (face-centered-cubic) or molten FeS mineralogies described by \citet{cammarano2006longperiod} (Table~\ref{table:Fe}). 
The density of the solid core for $X_{FeS}<5$\% is
\begin{equation}
\rho_{core}=\frac{\rho_{Fe}\rho_{FeS}}{X_{FeS}(\rho{Fe}-\rho_{FeS})+\rho_{Fe}}
\end{equation}
The choice of core composition does not affect the results for the properties of the deep ocean, which is the main focus of this paper. 

\begin{table}[h]
\caption{Iron and Iron-Sulfur properties.}
\begin{tabular}{|c|c|c|}
\hline
  & fcc-$\gamma$-Iron$^a$ & Fe-S (5\%$<X_{FeS}<20$\%)$^b$ \\
\hline
  $\rho$						&	8000	&	5150-(\%S-10)$\times$50				\\
$\alpha$,~10$^{-5}$~K$^{-1}$	&	5		&	9.2			\\
 $K_S$,~GPa						&	156		&	53.2-(\%S-10)$\times$2				\\
$\partial K_S/\partial P$		&	5.0		&	4.66			\\
 $\partial K_S/\partial T$,~Pa~K$^{-1}$	&	-0.040	&	--			\\
$G$,~GPA						&	 76.5	 &	--	     	\\
$\partial G/\partial P$ 		&    2   	&   --          \\
$\partial G/\partial T$,~Pa~K$^{-1}$ &  -0.023     &  --             \\
\hline
\multicolumn{3}{l}{$^a$\citet{cammarano2006longperiod};$^b$\citet{sanloup2000density}}
\end{tabular}
\label{table:Fe}
\end{table}

\subsection{Anelasticity (Seismic Attenuation)}

Frequency dependent seismic attenuation is computed as \citep{cammarano2006longperiod}
\begin{equation}
\frac{Q_S}{\omega^\gamma}=B_a\exp\bigg(\frac{\gamma H(P)}{RT}\bigg), H(P)=g_a T_m
\end{equation}
in which $B_a=0.56$ is a normalization factor, $\omega$ is the seismic frequency, exponent $\gamma=0.2$ is the frequency dependence of attenuation. $H$, the activation enthalpy, scales with the melting temperature $T_m$ with the anisotropy coefficient $g_a$. Chosen values of $g_a$ for ices \{I,II,III,V,VI\} are \{22, 30 25, 27, 28, 30\}. For the rocky mantle we assume $g_a$=30. Attenuation in Earth's ocean is small ($Q>10^4$), and is ignored.  If a metallic core is present, its value of $Q_S/\omega^{\gamma}$ is set to $10^4$.


\section{Results}\label{section:results}
 
For each ocean world, we display the bulk interior structure, and detailed ocean and ice structure for adiabatic profiles spanning the  range of  heat flux values consistent with the inferred existence of an ocean.

\subsection{Ganymede}
\begin{figure*} [h]
\noindent\includegraphics[width=35pc]{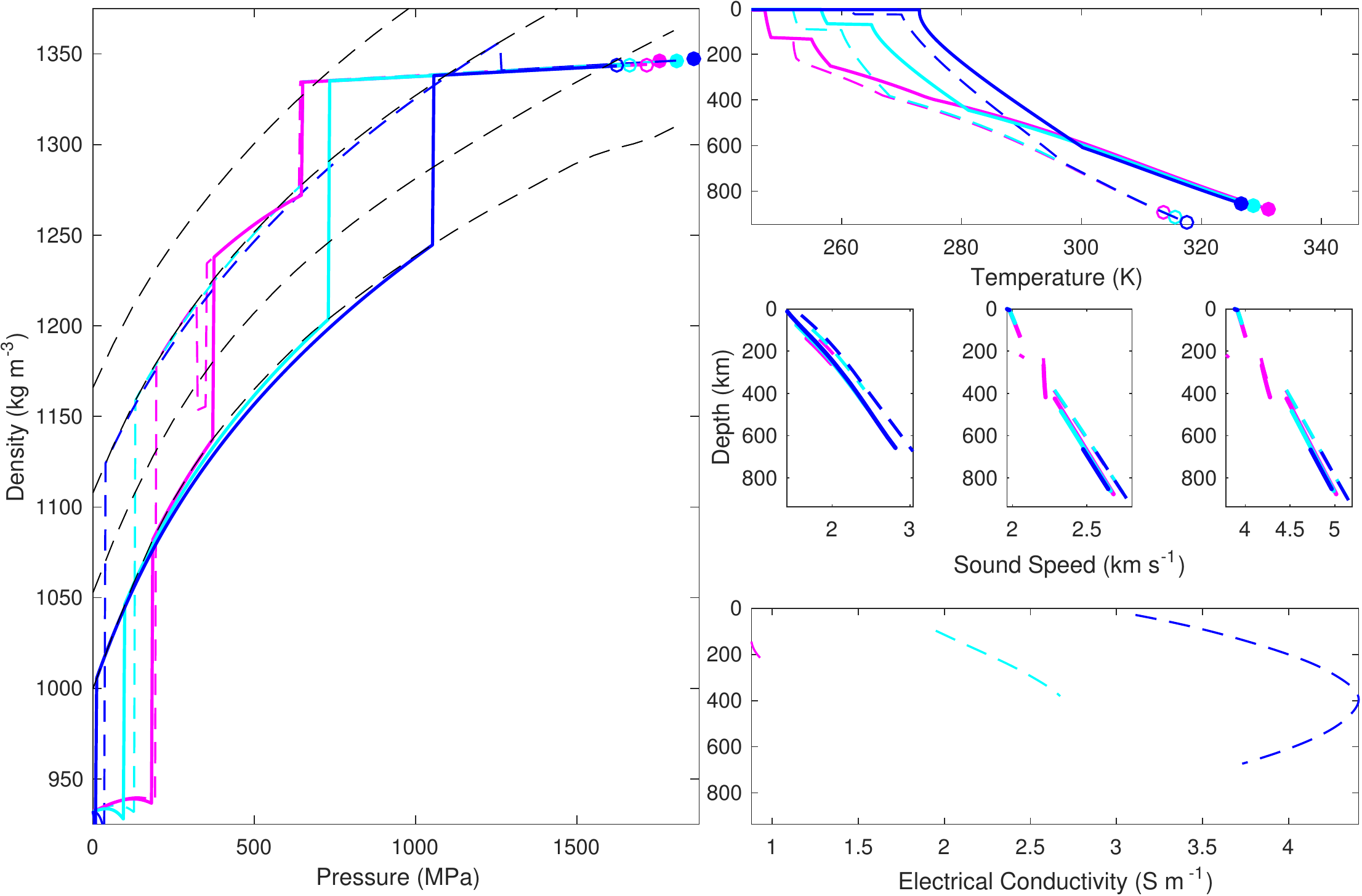}
\caption{Ganymede: Oceans with  10~wt\% MgSO$_4$(aq) (--), pure water (-~-) for $T_b$ as in Table~\ref{table:thicknessesGanymede}. Left: Density versus pressure. Reference fluid densities along the melting curve for MgSO$_4$ for \{0,5,10,15\}~wt\% increase with increasing concentration. Right: Corresponding depth-dependent temperature (top),  sound speed in the fluids and ices (fluid, $V_S$, $V_P$ middle left to right), and electrical conductivity (bottom). Circles indicate the transition to the silicate interior.}
\label{figure:ganymedeProfile10}
\end{figure*}

The structure for Ganymede is nearly identical to those discussed by \citet{vance2014ganymede}, but with the addition of solid state convection in the ice~I layer. Figure~\ref{figure:ganymedeProfile10} shows the configuration of density as a function of pressure, and corresponding temperature, sound speed, and electrical conductivity versus inferred depth for oceans with 0 and 10~wt\% MgSO$_4$. For the chosen ocean salinity and core composition (Table~\ref{table:thicknessesGanymede}), the silicate layer depth exceeds 800~km. Buoyant ice~III occurs at the base of the ocean for the coldest adiabat. For the warmest adiabat, ice~VI is also buoyant and ice~V is absent.    

Heat flux spans a range from 15 to 25~mW$^{-2}$, consistent with model-based constraints on Ganymede's thermal-orbital history \citep[e.g.,][]{bland2009orbital}. Higher heat fluxes, and ice as thin as $\sim$10~km are still in accord with the formation of grooved terrains on the surface \citep{hammond2014formation}. However these would require tidal heating. At present, such heating is probably negligible \citep[e.g.,][]{bland2015forming}, but residual heat may also lead to unexpectedly thinner ice than considered here.  Solid state convection is nearly four times more efficient than conductive cooling for the coldest adiabat.   

The convecting ocean shows temperature increasing strongly with depth due to the high compressibility of the fluids at 100~MPa relative to the solids. The addition of MgSO$_4$ decreases the compressibility and steepens the adiabat. Sound speeds are distinct for the different ice phases, and also for the freshwater and saline oceans. Sound speeds increase with depth everywhere. 

The temperature dependence of electrical conductivity  distinguishes among the different profiles.  The two warmest adiabats extend to sufficiently high pressure that the pressure derivative of electrical conductivity changes from increasing to decreasing. This is in accord with recent predictions investigations of water's interaction with dissolved ions at high pressures \citep{vance2015thermodynamics,schmidt2017pressure}.

\begin{figure*}[h]
\noindent\includegraphics[width=35pc]{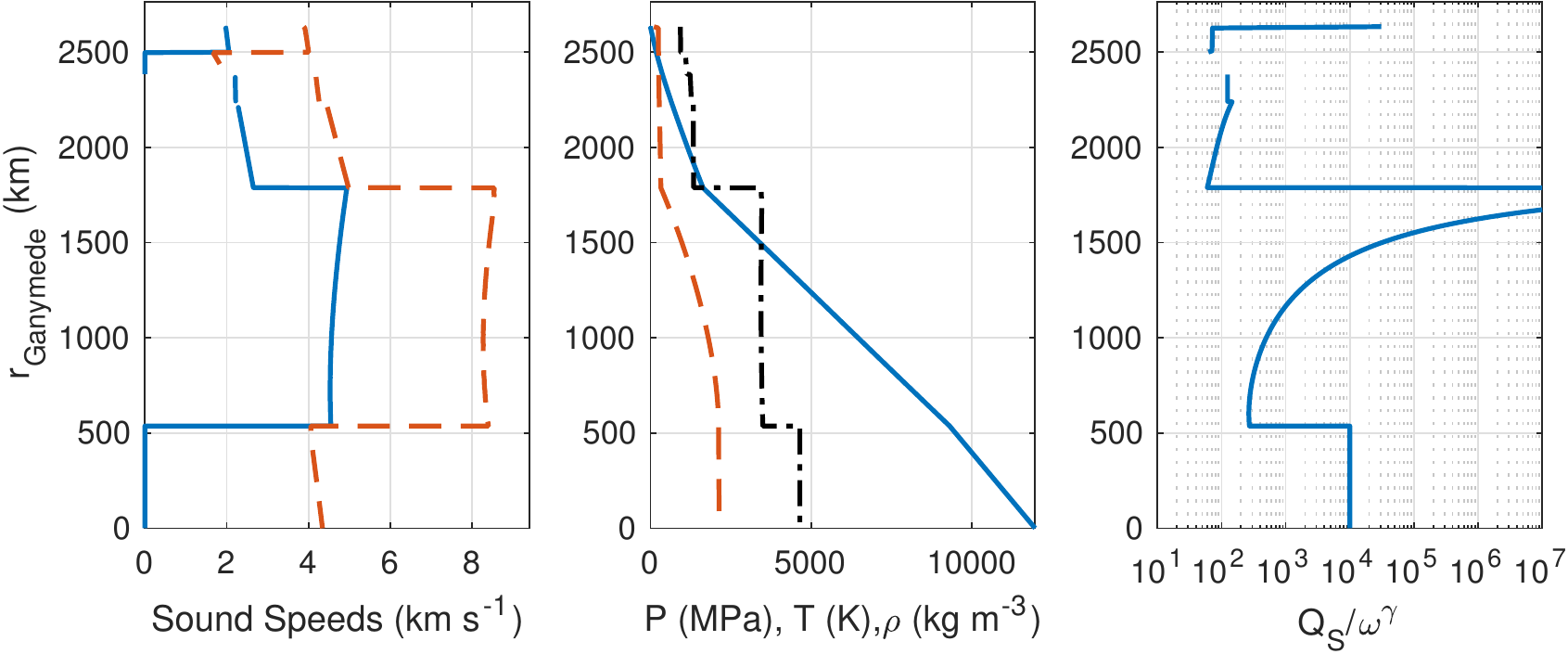}
\caption{Ganymede: global interior structure for an ocean with 10~wt\% MgSO$_4$(aq) and ice Ih thickness of 134~km. Left: $V_S$ (---) and $V_P$ (-- --). Middle:  Temperature (---) and density (-- --). Right: Anelasticity.}
\label{figure:ganymedeAttenuation10}
\end{figure*}

Figure~\ref{figure:ganymedeAttenuation10} shows the global structure: P- and S- wave speeds, pressure temperature and density, and anelasticity. The best fit mineral structure is a chondrite with 1wt\% hydration; Table~\ref{table:chondrite}). The presence of a liquid core is well documented for Ganymede \citep{kivelson2002permanent}. We specify a 20~wt\% FeS core composition.



\begin{table}[h]
\begin{center}
\begin{tabular}{|l|l|c|c|c|}
\hline
MgSO$_{4}$&$\rho_{rock}$ (kg m$^{-3}$)&\multicolumn{3}{c|}{3450}\\
\cline{3-5}
10wt\%&$\rho_{rock,model}$ (kg m$^{-3}$) &3457 &3456 &3456\\
&T$_{b}$ (K)   &252 &260 &270 \\
&q$_{b}$ mW m$^{-2}$   &4 &6 &22 \\
&q$_{c}$ mW m$^{-2}$   &15 &18 &25 \\
&$D_{Ih}$ (km) &142 &95 &26 \\
&$D_{ocean}$ (km) &75 &288 &650 \\
&$D_{III}$ (km) &16& -& - \\
&$D_{V}$ (km) &152& -& - \\
&$D_{VI}$ (km) &479 &501 &231 \\
&$R_{rock}$ (km) &1770 &1750 &1727 \\
&R$_{core}$ (km) &589 &629 &680 \\
\hline
\hline
Water&$\rho_{rock}$ (kg m$^{-3}$)&\multicolumn{3}{c|}{3450}\\
\cline{3-5}
&$\rho_{rock,model}$ (kg m$^{-3}$) &3456 &3457 &3458\\
&T$_{b}$ (K)   &255 &265 &273 \\
&q$_{b}$ mW m$^{-2}$   &4 &8 &107 \\
&q$_{c}$ mW m$^{-2}$   &16 &20 &107 \\
&$D_{Ih}$ (km) &134 &70 &5 \\
&$D_{ocean}$ (km) &117 &375 &603 \\
&$D_{V}$ (km) &144& -& - \\
&$D_{VI}$ (km) &452 &386 &215 \\
&$R_{rock}$ (km) &1787 &1803 &1811 \\
&R$_{core}$ (km) &537 &490 &479 \\
\hline
\end{tabular}
\end{center}
\caption{Ganymede: chondrite composition with 1~wt\% hydration.  $Q_{rock}=400$~GW.}
\label{table:thicknessesGanymede}
\end{table}%

\subsection{Europa}

\begin{figure*}[t]
\noindent\includegraphics[width=35pc]{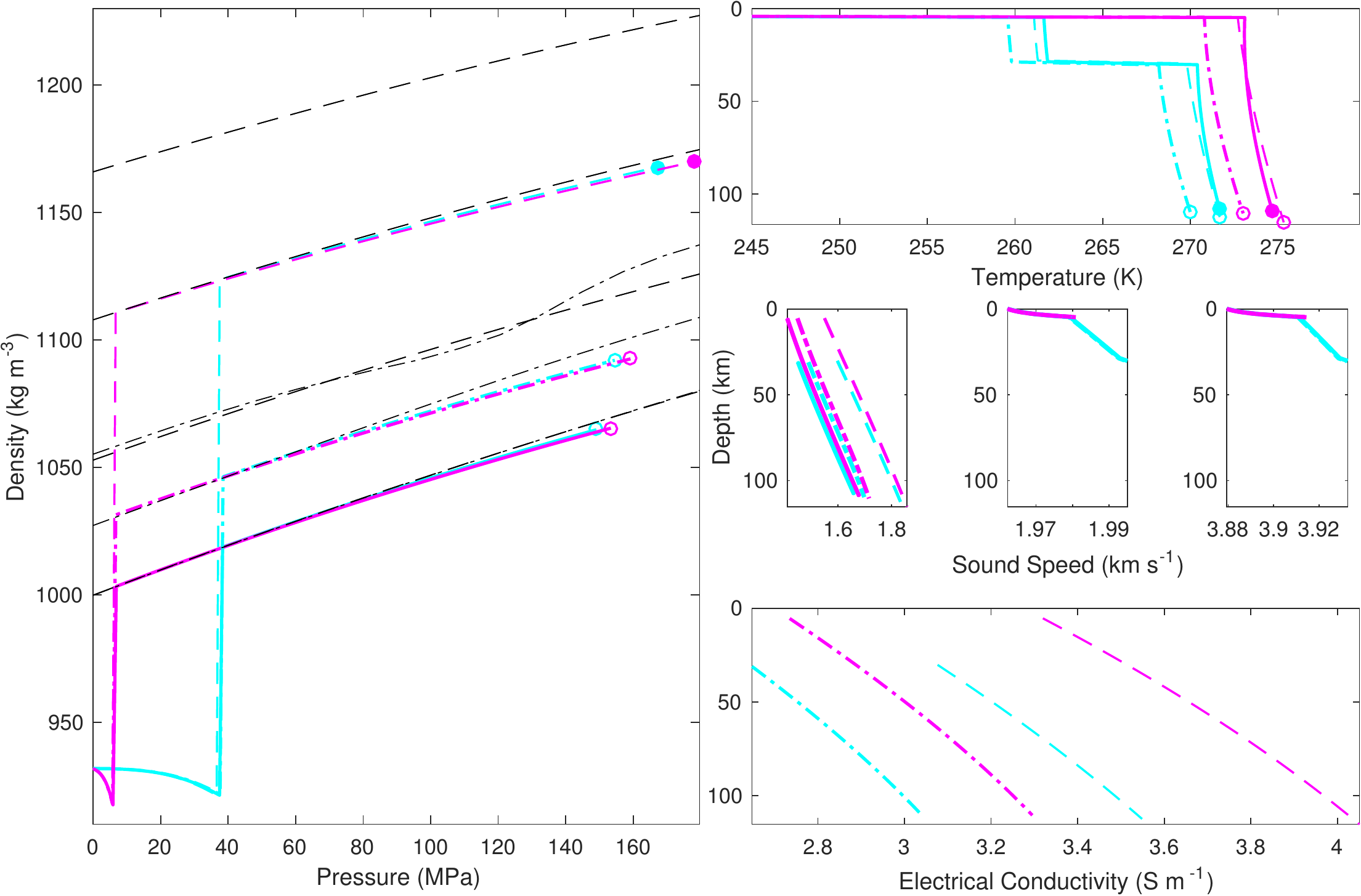}
\caption{Europa: Oceans with  10~wt\% MgSO$_4$(aq) (--), pure water (-~-), and standard seawater ($S_0$=35.165 g per kg fluid; --~$\cdot$) for $T_b$ as in Table~\ref{table:thicknessesEuropa}. Left: Density versus pressure. Reference fluid densities along the melting curve for MgSO$_4$ are \{0,5,10,15\}~wt\% and for seawater are \{0,1,2\}$\times S_0$. Right: Corresponding depth-dependent temperature (top),  sound speed in the fluids and ices (fluid, $V_S$, $V_P$ middle left to right), and electrical conductivity (bottom). Circles indicate the transition to the silicate interior.}
\label{figure:europaProfile}
\end{figure*}

Models for Europa (Figs.~\ref{figure:europaProfile} and \ref{figure:europaAttenuation10}) are adjusted through the choice of $T_b$ to compare ice thicknesses of 5 and 30~km. Model europas have seafloor depths slightly in excess of 100~km, consistent with the choice of silicate density and the assumption of a metallic core. 

For identical ice thicknesses, seawater salinity suppresses the melting of ice more strongly than MgSO$_4$. Sound speeds for seawater are comparable to those of water, whereas MgSO$_4$ sound speeds are larger by more than 10\%.  Oceans with MgSO$_4$ have higher electrical conductivity due to both higher ionic strength and higher temperature. Reference density profiles along the melting curve for seawater are unstable above 80~MPa for concentrations twice that of seawater (2$S_0$).

Sound speeds are low in the hydrous upper portion of the rocky interior, increasing with depth due to dehydration. The Na$_2$O-bearing pyrolite with 1~wt\% hydration provides an average rock density closest to the input value of 3300~kg~m$^{-3}$. We specify a core composition of pure iron.

\begin{table}[h]
\begin{center}
\begin{tabular}{|l|l|c|c|}
\hline
MgSO$_{4}$&$\rho_{rock}$ (kg m$^{-3}$)&\multicolumn{2}{c|}{3300}\\
\cline{3-4}
10wt\%&$\rho_{rock,model}$ (kg m$^{-3}$) &3375 &3375\\
&T$_{b}$ (K)   &269.80 &272.70 \\
&q$_{b}$ mW m$^{-2}$   &19 &123 \\
&q$_{c}$ mW m$^{-2}$   &24 &123 \\
&$D_{Ih}$ (km) &30 &5 \\
&$D_{ocean}$ (km) &88 &116 \\
&$R_{rock}$ (km) &1443 &1440 \\
&R$_{core}$ (km) &522 &523 \\
\hline
\hline
Water&$\rho_{rock}$ (kg m$^{-3}$)&\multicolumn{2}{c|}{3300}\\
\cline{3-4}
&$\rho_{rock,model}$ (kg m$^{-3}$) &3375 &3374\\
&T$_{b}$ (K)   &270.40 &273.10 \\
&q$_{b}$ mW m$^{-2}$   &19 &119 \\
&q$_{c}$ mW m$^{-2}$   &24 &119 \\
&$D_{Ih}$ (km) &30 &5 \\
&$D_{ocean}$ (km) &83 &109 \\
&$R_{rock}$ (km) &1448 &1447 \\
&R$_{core}$ (km) &521 &521 \\
\hline
\hline
Seawater&$\rho_{rock}$ (kg m$^{-3}$)&\multicolumn{2}{c|}{3300}\\
\cline{3-4}
35.165g/kg&$\rho_{rock,model}$ (kg m$^{-3}$) &3375 &3375\\
&$X_{FeS}$ (\%)&\multicolumn{2}{c|}{0}\\
&T$_{b}$ (K)   &268.20 &270.80 \\
&q$_{b}$ mW m$^{-2}$   &18 &121 \\
&q$_{c}$ mW m$^{-2}$   &23 &121 \\
&$D_{Ih}$ (km) &30 &5 \\
&$D_{ocean}$ (km) &84 &111 \\
&$R_{rock}$ (km) &1447 &1446 \\
&R$_{core}$ (km) &521 &520 \\
\hline
\end{tabular}

\end{center}
\caption{Europa:pyrolite composition with Na$_2$O. $Q_{rock}=100$~GW.}
\label{table:thicknessesEuropa}
\end{table}%

\begin{figure*}[h]
\noindent\includegraphics[width=35pc]{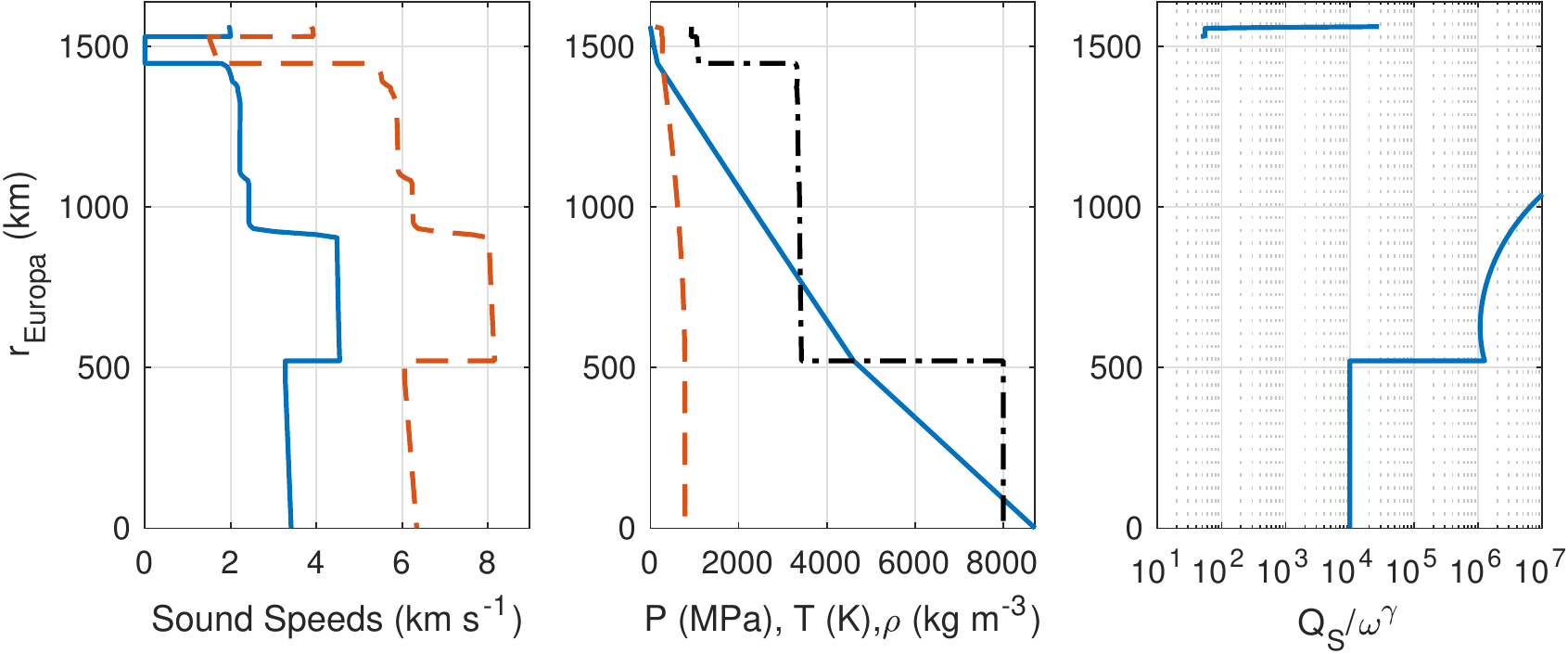}
\caption{Europa: global interior structure for an ocean with the composition seawater and 30\-km-thick ice Ih lithosphere. Left: $V_S$ (---) and $V_P$ (-- --). Middle:  Pressure (---), temperature (--~--), and density ($\cdot$~--). Right: Anelasticity.}
\label{figure:europaAttenuation10}
\end{figure*}

\subsection{Enceladus}

\begin{figure*}[h]
\noindent\includegraphics[width=35pc]{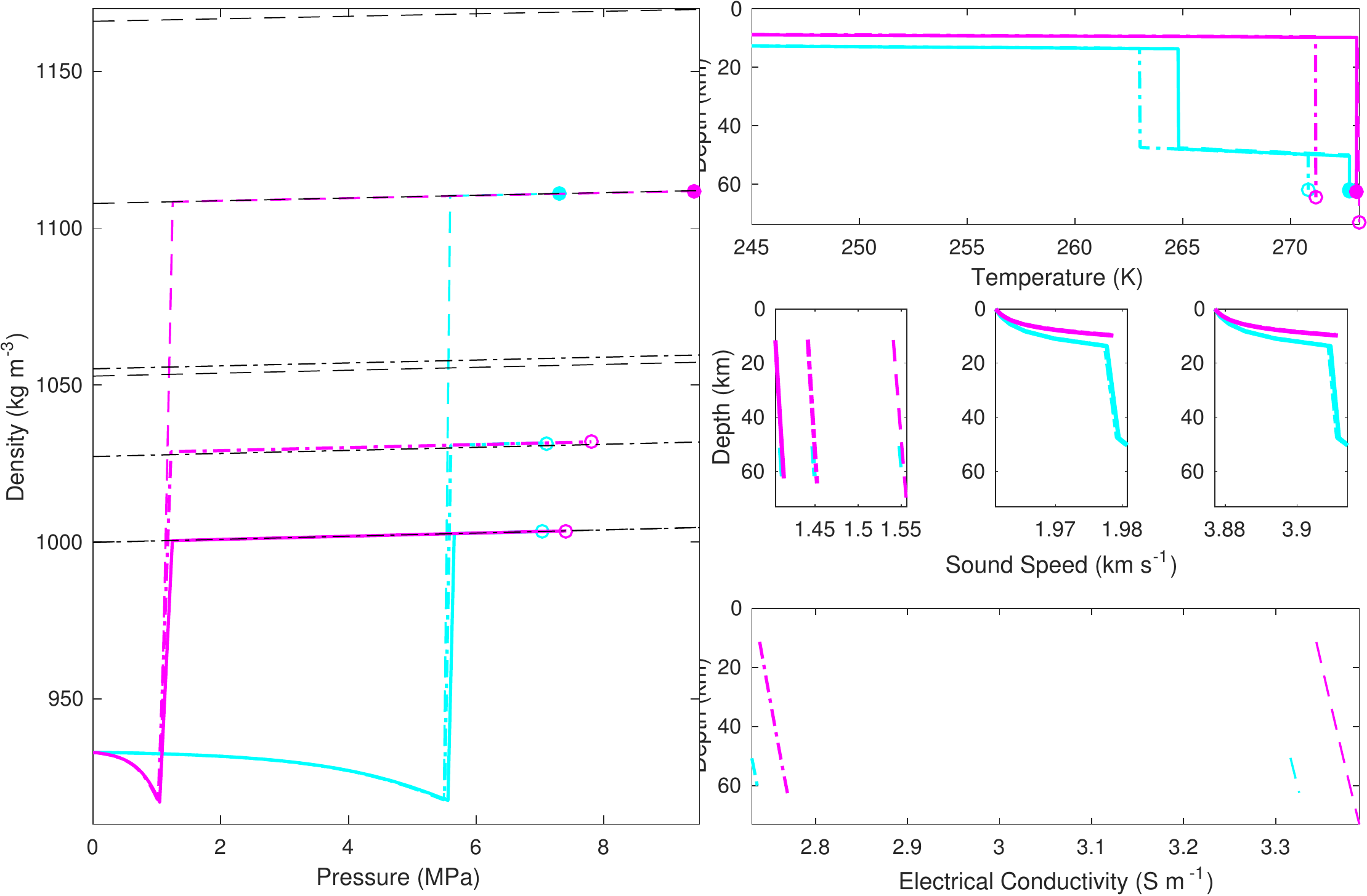}
\caption{Enceladus:  Oceans with  10~wt\% MgSO$_4$(aq) (--), pure water (-~-), and standard seawater (--~$\cdot$) for $T_b$ as in Table~\ref{table:thicknessesEnceladusHydrousPyrolite}. Left: Density versus pressure. Reference densities as for Europa (Fig.~\ref{figure:europaProfile}. Right: Corresponding depth-dependent temperature (top),  sound speed in the fluids and ices (fluid, $V_S$, $V_P$ middle left to right), and electrical conductivity (bottom). Circles indicate the transition to the silicate interior.}
\label{figure:enceladusProfile}
\end{figure*}

Models for Enceladus (Figs. \ref{figure:enceladusProfile}) are adjusted through the choice of $T_b$ to compare ice thicknesses of 10 and 50~km corresponding to constraints at the south polar plume and in the rest of the global ocean, respectively \citep{mckinnon2015effect,beuthe2016crustal,thomas2016enceladus}. MgSO$_4$ is considered, despite strong inferences that the ocean has a high pH and thus is dominated by chlorides \citep{postberg2011salt,glein2015ph}. As with Europa, the different ocean compositions have distinct profiles in temperature, density, electrical conductivity, and sound speed.  For models examining a 10-km-thick ice~Ih layer potentially present in the south polar region, the depth of the silicate interior varies by up to 10~km with salinity due to the large proportional change in density of the volatile layer.

The small size of Enceladus precludes the existence of a metallic core. All models matching the bulk ice thickness of 50~km require a low rock density ($\approx2700$~kg~m$^{-3}$).  This is met by either a fully hydrous pyrolite (Table \ref{table:thicknessesEnceladusHydrousPyrolite}) or anhydrous and porous chondrite
. The heat flux in the rocky interior is set to 0.27~GW. As shown in Fig.~\ref{figure:enceladusPorosity}, porosity introduces a strong temperature dependence to density and sound speeds, reducing the density by up to more than 150\% and sound velocity by up to more than 75\%. If the 10~km ice thickness were the global value, the smaller rock density ($\approx2200$~kg~m$^{-3}$) could be met by adding a combination of Na$_2$O, hydration, and porosity. 
The anelasticity of the rocky layer is high ($Q_{S} / \omega^\gamma>10^7$) in all cases.

\begin{table}[h]
\begin{center}
\begin{tabular}{|l|l|c|c|c|c|}
\hline
MgSO$_{4}$&$\rho_{rock}$ (kg m$^{-3}$) &2701 &2265\\
10wt\%&$\rho_{rock,model}$ (kg m$^{-3}$) &2698 &2697\\
&T$_{b}$ (K)   &272.72 &273.12 \\
&q$_{b}$ mW m$^{-2}$   &16 &83 \\
&q$_{c}$ mW m$^{-2}$   &10 &83 \\
&$D_{Ih}$ (km) &50 &10 \\
&$D_{ocean}$ (km) &13 &63 \\
&$R_{rock}$ (km) &190 &179 \\
\hline
\hline
Seawater&$\rho_{rock}$ (kg m$^{-3}$) &2669 &2222\\
35.165g/kg&$\rho_{rock,model}$ (kg m$^{-3}$) &2698 &2698\\
&T$_{b}$ (K)   &270.82 &271.16 \\
&q$_{b}$ mW m$^{-2}$   &16 &84 \\
&q$_{c}$ mW m$^{-2}$   &10 &84 \\
&$D_{Ih}$ (km) &50 &10 \\
&$D_{ocean}$ (km) &12 &55 \\
&$R_{rock}$ (km) &190 &188 \\
\hline
\hline
Water&$\rho_{rock}$ (kg m$^{-3}$) &2668 &2216\\
&$\rho_{rock,model}$ (kg m$^{-3}$) &2698 &2697\\
&T$_{b}$ (K)   &272.74 &273.08 \\
&q$_{b}$ mW m$^{-2}$   &16 &83 \\
&q$_{c}$ mW m$^{-2}$   &10 &83 \\
&$D_{Ih}$ (km) &50 &10 \\
&$D_{ocean}$ (km) &11 &53 \\
&$R_{rock}$ (km) &190 &189 \\
\hline
\hline
NH$_{3}$&$\rho_{rock}$ (kg m$^{-3}$) &2651 &2210\\
3wt\%&$\rho_{rock,model}$ (kg m$^{-3}$) &2698 &2698\\
&T$_{b}$ (K)   &269.54 &269.90 \\
&q$_{b}$ mW m$^{-2}$   &16 &81 \\
&q$_{c}$ mW m$^{-2}$   &10 &81 \\
&$D_{Ih}$ (km) &50 &10 \\
&$D_{ocean}$ (km) &12 &51 \\
&$R_{rock}$ (km) &191 &191 \\
\hline
\end{tabular}
\end{center}
\caption{Enceladus: saturated pyrolite composition. $Q_{rock}=0.27$~GW.}
\label{table:thicknessesEnceladusHydrousPyrolite}
\end{table}%

\begin{figure*}[h]
\noindent\includegraphics[width=35pc]{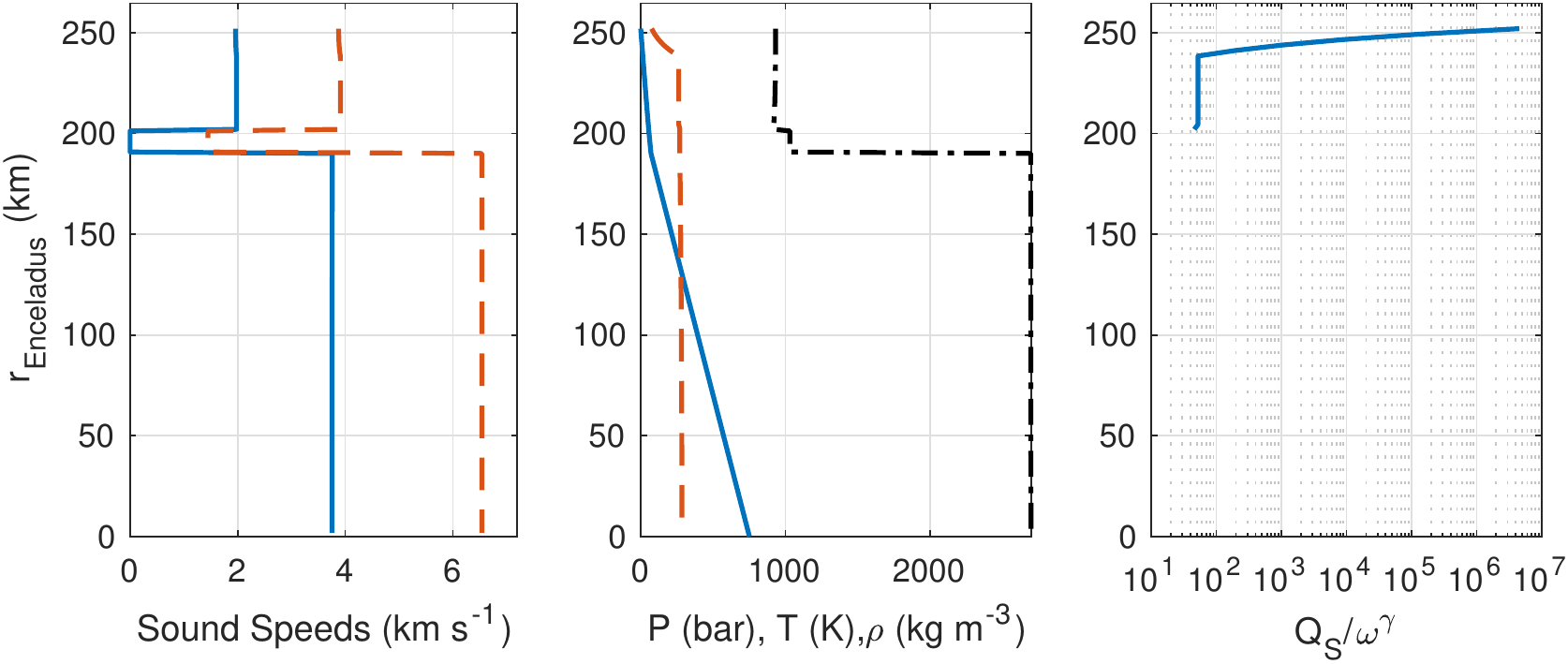}
\caption{Enceladus: global interior structure with the ocean composition of seawater. Left: $V_S$ (---) and $V_P$ (-- --). Middle:  Pressure (---), temperature (--~--), and density ($\cdot$~--). Right: Anelasticity.}
\label{figure:enceladusAttenuation}
\end{figure*}

\begin{figure*}[h]
\noindent\includegraphics[width=35pc]{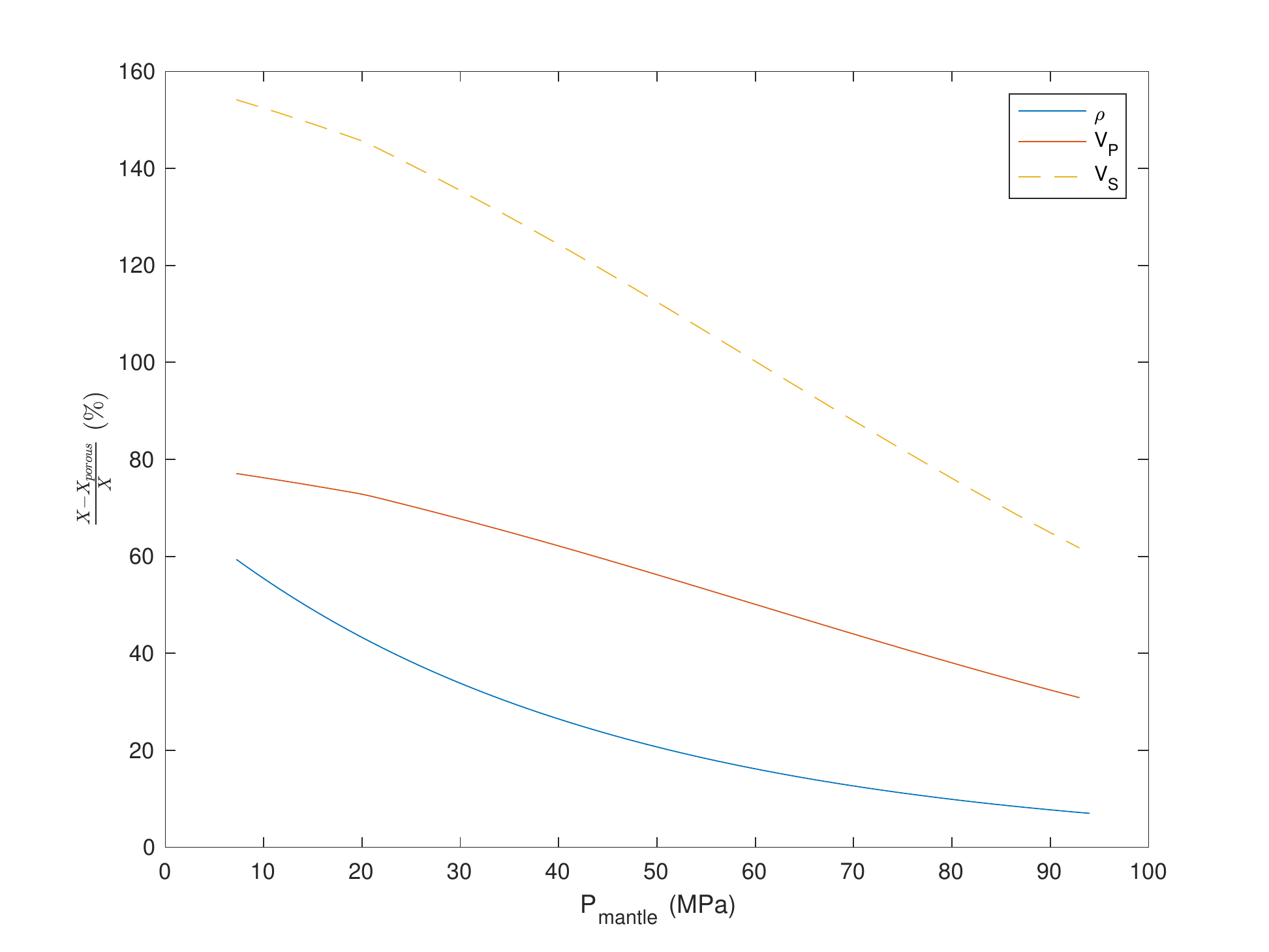}
\caption{Enceladus: The influence of porosity on density, $V_S$ (---) and $V_P$ (--~--) for an anhydrous chondrite composition.}
\label{figure:enceladusPorosity}
\end{figure*}

\subsection{Titan}
Titan's ice covering is interpreted to be 55-80~km based on the observed Schumann resonance \citep{beghin2012analytic}.  We consider thicknesses in this range and approaching as much as 150~km. 

We compare the effects of including small amounts of ammonia (3~wt\%) or requiring a high salinity of MgSO$_4$ (10~wt\%). The prevalence of reduced volatiles in Titan's atmosphere and seas suggests against the more oxidized sulfate-rich ocean studied by \citet{fortes2007ammonium,grindrod2008long}, but we consider sulfate due to the availability of the equation of state. 

\begin{figure*}[h]
\noindent\includegraphics[width=35pc]{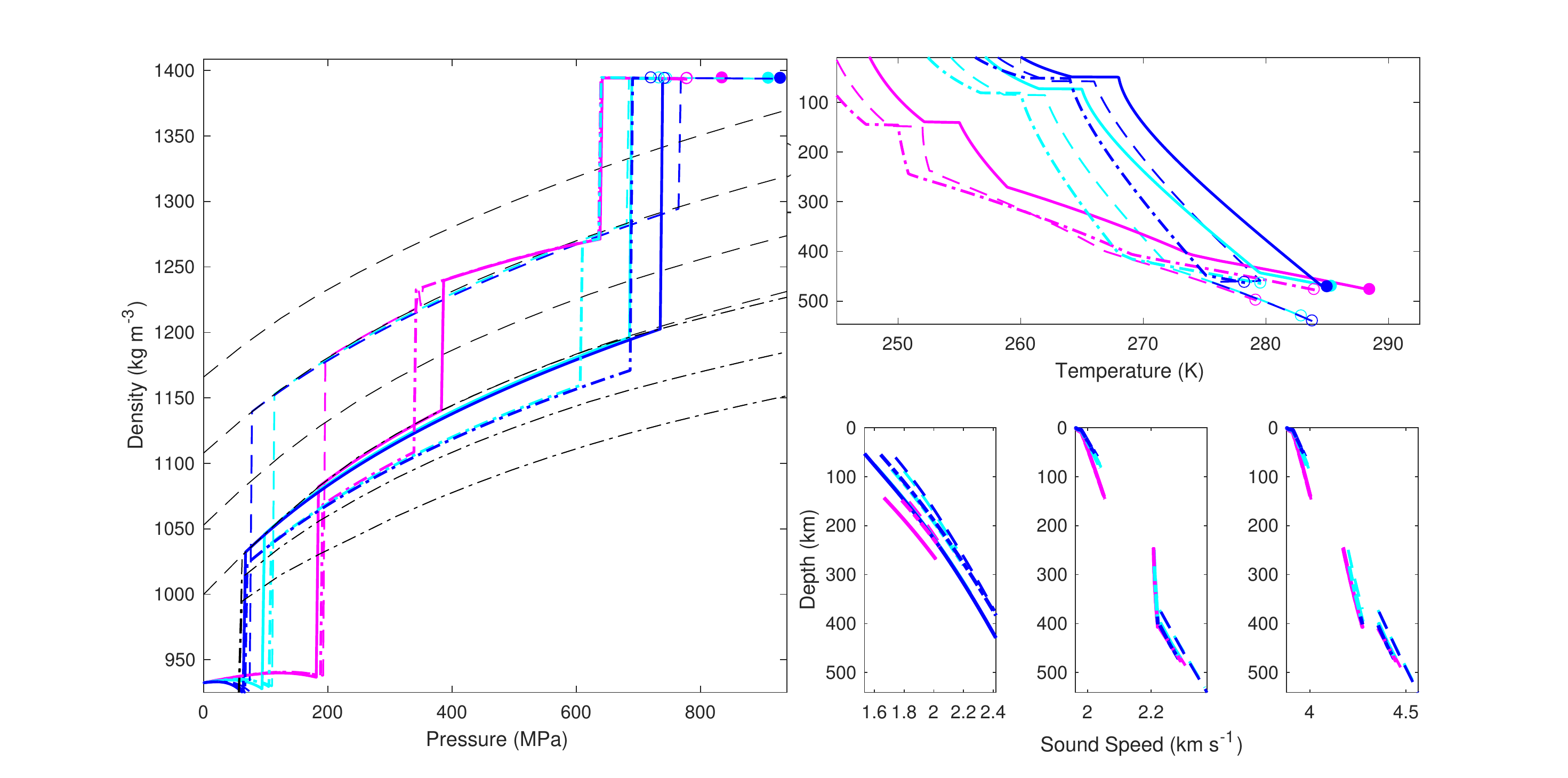}
\caption{Titan: Oceans with  3~wt\% NH$_3$(aq) (--~$\cdot$), pure water (\textbf{--}), and MgSO$_4$ (-- --) for $T_b$ as in Table~\ref{table:thicknessesTitan_NoCore}.  Left: Density versus pressure. Reference fluid densities at \{0,5,10\} wt\% for NH$_3$ decrease with increasing concentration, and increase for MgSO$_4$. Right: Corresponding depth-dependent temperature (top),  sound speeds in the fluids and ices (fluid, $V_{S}$, $V_{P}$ bottom left to right). Circles denote the transition to the silicate interior.}
\label{figure:titanStructureNoCore}
\end{figure*}

We consider models without a metallic core Figure~\ref{figure:titanStructureNoCore} and Table~\ref{table:thicknessesTitan_NoCore}.   Heat flux in the rocky interior is set to 120~GW. For the pure water ocean (solid lines), ice V and VI are present, except in the warmest cases. For the highest bottom melting temperature ($T_{b}$=270~K), the ocean has no high pressure ices at all. Adding 3~wt\%~NH$_3$ to Titan's ocean lowers the density and narrows the adiabat. The lower fluid compressibility corresponds to a higher sound speed.   For the highest heat flow model (thinnest ice~Ih), the ammmonia and water oceans have little or no high pressure ice. The higher densities of the MgSO$_4$ oceans means high-pressure ices are thicker.

The required rock densities are low ($\approx2600$~kg~m$^{-3}$), most consistent with the saturated chondrite composition.

\begin{table}[h]
\begin{center}
\begin{tabular}{|l|l|c|c|c|c|}
\hline
MgSO$_{4}$&$\rho_{rock}$ (kg m$^{-3}$) &2616 &2545 &2524\\
10wt\%&$\rho_{rock,model}$ (kg m$^{-3}$) &2623 &2562 &2468\\
\cline{3-5}
&$Q_{rock}$ (GW)&\multicolumn{3}{c|}{120   }\\
\cline{3-5}
&T$_{b}$ (K)   &252 &262 &266 \\
&q$_{b}$ mW m$^{-2}$   &4 &8 &11 \\
&q$_{c}$ mW m$^{-2}$   &14 &17 &19 \\
&$D_{Ih}$ (km) &149 &86 &58 \\
&$D_{ocean}$ (km) &91 &333 &403 \\
&$D_{V}$ (km) &163& -& - \\
&$D_{VI}$ (km) &96 &111 &79 \\
&$R_{rock}$ (km) &2076 &2044 &2034 \\
\hline
\hline
Water&$\rho_{rock}$ (kg m$^{-3}$) &2540 &2527 &2517\\
&$\rho_{rock,model}$ (kg m$^{-3}$) &2611 &2598 &2591\\
\cline{3-5}
&$Q_{rock}$ (GW)&\multicolumn{3}{c|}{120       }\\
\cline{3-5}
&T$_{b}$ (K)   &255 &265 &268 \\
&q$_{b}$ mW m$^{-2}$   &4 &9 &13 \\
&q$_{c}$ mW m$^{-2}$   &14 &18 &20 \\
&$D_{Ih}$ (km) &141 &74 &50 \\
&$D_{ocean}$ (km) &130 &369 &420 \\
&$D_{V}$ (km) &140& -& - \\
&$D_{VI}$ (km) &67 &27 &2 \\
&$R_{rock}$ (km) &2097 &2104 &2103 \\
\hline
\hline
NH$_{3}$&$\rho_{rock}$ (kg m$^{-3}$) &2531 &2523 &2519\\
3wt\%&$\rho_{rock,model}$ (kg m$^{-3}$) &2612 &2603 &2601\\
\cline{3-5}
&$Q_{rock}$ (GW)&\multicolumn{3}{c|}{120         }\\
\cline{3-5}
&T$_{b}$ (K)   &250 &260 &264 \\
&q$_{b}$ mW m$^{-2}$   &4 &8 &12 \\
&q$_{c}$ mW m$^{-2}$   &13 &17 &19 \\
&$D_{Ih}$ (km) &146 &82 &52 \\
&$D_{ocean}$ (km) &100 &321 &396 \\
&$D_{V}$ (km) &164 &16& - \\
&$D_{VI}$ (km) &68 &46 &15 \\
&$R_{rock}$ (km) &2097 &2110 &2112 \\
\hline
\end{tabular}
\end{center}
\caption{Titan: saturated chondrite composition. $Q_{rock}=120$~GW.}
\label{table:thicknessesTitan_NoCore}
\end{table}%

\begin{figure*}[h]
\noindent\includegraphics[width=35pc]{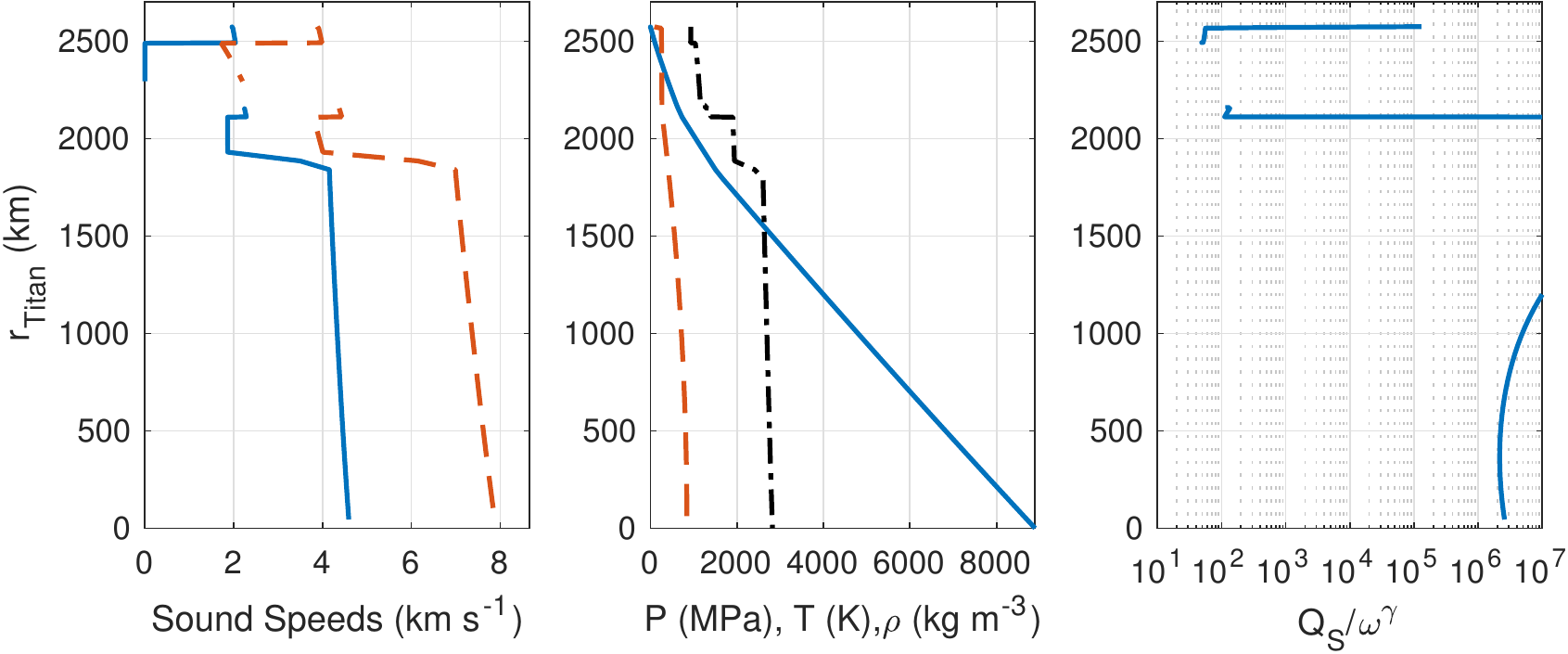}
\caption{Titan: global interior structure for an ocean with 3~wt\% NH$_3$(aq). Left: $V_S$ (---) and $V_P$ (-- --). Middle: Pressure (---), temperature (--~--), and density ($\cdot$~--). Right: Anelasticity.}
\label{figure:europaAttenuation10}
\end{figure*}

\subsection{Callisto}

We consider ice thicknesses exceeding 100~km. For simplicity we do not modify the parameterization of solid state convection to accommodate a fully stagnant lid \citep{mckinnon2006convection}. We consider oceans with 0~and 10~wt\% MgSO$_4$. 

\begin{figure*}[h]
\noindent\includegraphics[width=35pc]{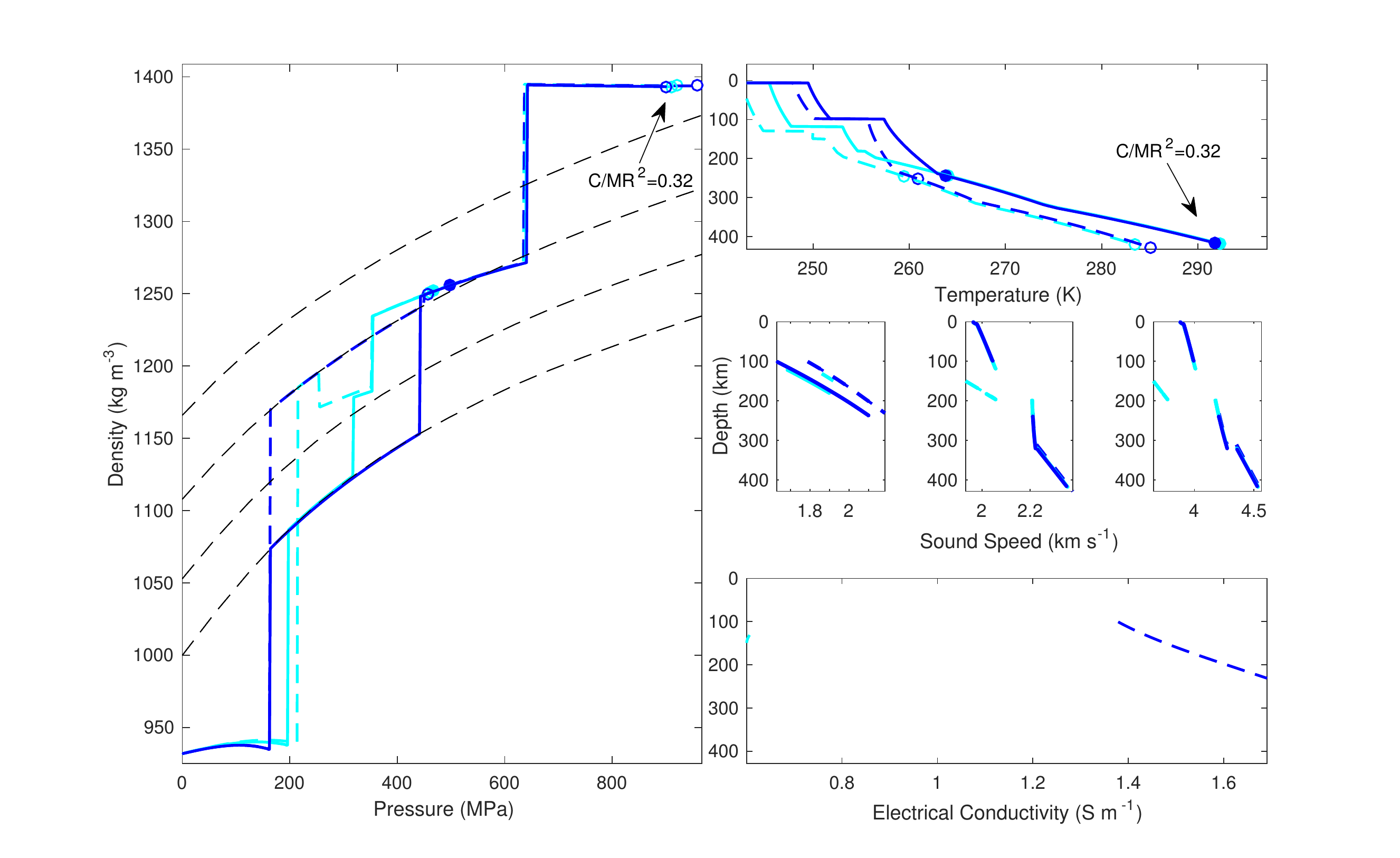}
\caption{Callisto:  Oceans with  10~wt\% MgSO$_4$(aq) (--), pure water (-~-), and standard seawater (--~$\cdot$) for $T_b$ as in Table~\ref{table:thicknessesCallisto}. Left: Density versus pressure. Reference densities as for Europa (Fig.~\ref{figure:europaProfile}). Right: Corresponding depth-dependent temperature (top),  sound speed in the fluids and ices (fluid, $V_S$, $V_P$ middle left to right), and electrical conductivity (bottom). Circles indicate the transition to the silicate interior. Greater silicate depths and pressures are also shown for the assumption of C/MR$^2$=0.32}
\label{figure:callistoStructure}
\end{figure*}

Figure~\ref{figure:callistoStructure} shows interiors for both the reported value of Callisto's gravitational moment of inertia \citep[$C/MR^2=0.3549\pm0.0042$;][Fig.~\ref{figure:callistoStructure}]{schubert2004interior} and a value that is  10\% smaller \citep[ $C/MR^2=0.32$;][]{gao2013nonhydrostatic}. The higher value of $C/MR^2$ precludes the presence of a metallic core and requires a low silicate density around 3100~kg~m$^{-3}$  (Table~\ref{table:thicknessesCallisto}), most consistent with the saturated pyrolite composition. Rock-interface depths of less than $\sim$250~km are required. Ice VI does not occur. Buoyant ice III is present for the coldest oceans containing both pure water and salt.
The smaller value of  $C/MR^2$ leads to silicate depths similar to Ganymede's (Table~\ref{table:thicknessesCallisto_SmallerCMR2} and the likelihood of an iron core, specified as pure ($\gamma$) iron. Rock densities are best fit by the anhydrous pyrolite composition. Ice~VI is always present, and buoyant ice III is present in the coldest model that includes salts. 


\begin{figure*}[h]
\noindent\includegraphics[width=35pc]{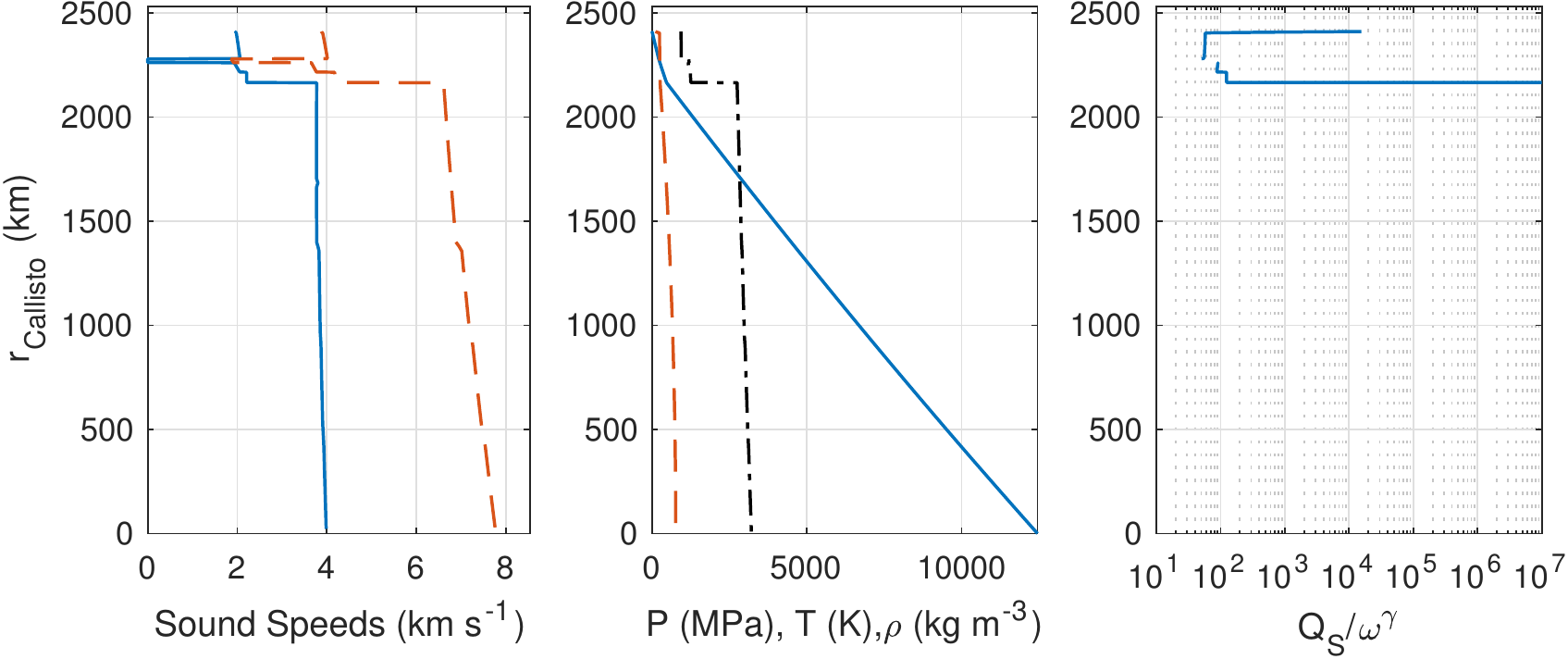}
\caption{Callisto: global interior structure for an ocean with 10~wt\% MgSO$_4$(aq). Left: $V_S$ (---) and $V_P$ (-- --). Middle:  Pressure (---), temperature (--~--), and density ($\cdot$~--). Right: Anelasticity.}
\label{figure:callistoAttenuation10}
\end{figure*}

\begin{table}[h]
\begin{center}
\begin{tabular}{|l|l|c|c|c|}
\hline
MgSO$_{4}$&$\rho_{rock}$ (kg m$^{-3}$) &3118 &3054\\
10wt\%&$\rho_{rock,model}$ (kg m$^{-3}$) &2933 &2935\\
&T$_{b}$ (K)   &250 &255.70 \\
&q$_{b}$ mW m$^{-2}$   &4 &5 \\
&q$_{c}$ mW m$^{-2}$   &16 &18 \\
&$D_{Ih}$ (km) &130 &100 \\
&$D_{ocean}$ (km) &20 &132 \\
&$D_{III}$ (km) &46& - \\
&$D_{V}$ (km) &120 &79 \\
&$R_{rock}$ (km) &2165 &2158 \\
\hline
\hline
Water&$\rho_{rock}$ (kg m$^{-3}$) &3094 &3077\\
&$\rho_{rock,model}$ (kg m$^{-3}$) &2936 &2937\\
&T$_{b}$ (K)   &253.10 &257.40 \\
&q$_{b}$ mW m$^{-2}$   &4 &5 \\
&q$_{c}$ mW m$^{-2}$   &17 &18 \\
&$D_{Ih}$ (km) &120 &100 \\
&$D_{ocean}$ (km) &62 &139 \\
&$D_{III}$ (km) &16& - \\
&$D_{V}$ (km) &121 &83 \\
&$R_{rock}$ (km) &2166 &2166 \\
\hline
\end{tabular}
\end{center}
\caption{Callisto: saturated pyrolite composition. $Q_{rock}=110$~GW}
\label{table:thicknessesCallisto}
\end{table}%

\begin{table}[h]
\begin{center}
\begin{tabular}{|l|l|c|c|c|}
\hline
MgSO$_{4}$&$\rho_{rock}$ (kg m$^{-3}$)&\multicolumn{2}{c|}{3500}\\
\cline{3-4}
10wt\%&$\rho_{rock,model}$ (kg m$^{-3}$) &3545 &3545\\
&T$_{b}$ (K)   &250 &255.70 \\
&q$_{b}$ mW m$^{-2}$   &4 &5 \\
&q$_{c}$ mW m$^{-2}$   &16 &18 \\
&$D_{Ih}$ (km) &130 &100 \\
&$D_{ocean}$ (km) &20 &132 \\
&$D_{III}$ (km) &46& - \\
&$D_{V}$ (km) &120 &79 \\
&$D_{VI}$ (km) &120 &127 \\
&$R_{rock}$ (km) &1975 &1973 \\
&R$_{core}$ (km) &617 &598 \\
\hline
\hline
Water&$\rho_{rock}$ (kg m$^{-3}$)&\multicolumn{2}{c|}{3500}\\
\cline{3-4}
&$\rho_{rock,model}$ (kg m$^{-3}$) &3543 &3543\\
&T$_{b}$ (K)   &253.10 &257.40 \\
&q$_{b}$ mW m$^{-2}$   &4 &5 \\
&q$_{c}$ mW m$^{-2}$   &17 &18 \\
&$D_{Ih}$ (km) &120 &100 \\
&$D_{ocean}$ (km) &62 &139 \\
&$D_{III}$ (km) &16& - \\
&$D_{V}$ (km) &121 &83 \\
&$D_{VI}$ (km) &116 &112 \\
&$R_{rock}$ (km) &1975 &1976 \\
&R$_{core}$ (km) &623 &624 \\
\hline
\end{tabular}

\end{center}
\caption{Callisto with C/MR$^2$=0.32 and anhydrous pyrolite composition. $Q_{rock}=110$~GW}
\label{table:thicknessesCallisto_SmallerCMR2}
\end{table}%


\section{Discussion}\label{section:discussion}
The computed profiles demonstrate the possible use of geophysical measurements to assess the habitability of ocean worlds. The combination of ocean salinity, bulk density, density structure, and ice I thickness constitute a combined constraint on the properties of the constituent materials, which must be logically consistent. Heat fluxes---which have yet to be measured directly---couple with salinity to determine which crystalline forms of ice are present in what amounts, and whether they will sink or float. 

Consideration of the rocky layers of known ocean worlds suggests that a combination of measurements may reduce to a manageably short list or uniquely determine the possible combined properties of hydration state, porosity, and oxide content. For example, the oxide composition and hydration state create density signatures and phase boundaries that may be inferred from gravitational and seismic investigations, respectively. 

Water's main influence on density and seismic properties is to significantly reduce the solidus temperature.
For scenarios that include a hot rocky interior, this may affect the viscoelastic relaxation of rocks, also at seismic frequencies (as later discussed in the anelasticity section) and already at temperatures lower than the solidus.

\subsection{Tests for Habitability}
Depth-dependent density, temperature, electrical conductivity, and sound speed can be used to locate liquid reservoirs and thus confirm water-rock interactions. Inferring ocean composition and pH implies a constraint on the integrated redox fluxes through time. 

\subsubsection{Europa} 
The seawater and MgSO$_4$ oceans examined for Europa span the difference between oxidized and low-pH or reduced and high-pH, as simulated by \citet{zolotov2008oceanic}. For similar ice thicknesses, the higher-pH chloride ocean has a lower melting temperature, which dictates lower electrical conductivity, density, and sound speed. The first two of these will be constrained by gravity and electromagnetic experiments on the JUICE \citep{grasset2013JUICE} and Europa \citep{pappalardo2016science} missions in combination with radar sounding and topographic imaging. A subsequent seismic experiment could address any degeneracies in interpreted ocean density and electrical conductivity by measuring the oceanic sound speed. Thermal radar mapping will establish the existence and characteristics of solid state convection, providing additional prior constraints for a follow-on seismic investigation. With this background established, it may be possible to determine the extent of aqueous alteration and sedimentation at the seafloor \citep{vance2016seismic}.

\subsubsection{Enceladus} 
The silicate interior of Enceladus controls the extent of water rock interaction, including the production of hydrogen identified in south polar plume materials \citep{waite2017cassini}.  Thought it has previously suggested that the rocky core of Enceladus must be fully hydrated \citep{vance2007hydrothermal,vance2016geophysical} as per the hydrous pyrolite model used here, the low pressures at Enceladus also permit an anhydrous chondrite composition with high porosities consistent with those in Earth's upper mantle. A transition from anhydrous to hydrous would explain the apparent present-day production of hydrogen \citep{hsu2015ongoing}, and might be consistent with recent activation or even formation of Enceladus \citep{cuk2016dynamical}.

Electromagnetic and seismic measurements could evaluate the extent and role of fluids and gases in south polar plume eruptions \citep{postberg2011salt,hsu2015ongoing,kite2016sustained}. By analogy with terrestrial investigations of geysers and volcanoes, a seismic experiment could reveal the flow rates, compositions, and geometry of plumes on Enceladus (Vance et al., under revision; Stähler et al., submitted). The seismic detectability of the strong influence of rock porosity on sound speed and density should be investigated.

\subsubsection{Titan} 
A bulk ocean density of $>1200$~kg~m$^{-3}$ is required of the large value of the measured tidal Love number ($k_2 >0.6$) is correct \citet{mitri2014shape}. The 10~wt\%~MgSO$_4$ ocean meets this requirement. As noted above, a reducing ocean dominated instead by chlorides can obtain similarly large densities. A saline ocean is not required, however, as the uncertainty in the Love number ($k_2= 0.637\pm0.220$) permits densities as low as $1100$~kg~m$^{-3}$, consistent with the pure water or 3~wt\% cases studied here. A small core ($R<400$~km) coupled to a cold hydrous mantle, permitted by the gravitational constraints, would suggest an early hot Titan that lost most of its internal heat. A low heat flux in the deep interior is hard to reconcile with the inference that the ice is thinner than 100~km, unless the ice or ocean are tidally heated. 
If Titan's ice is indeed thin and if the ocean has a low salinity, high-pressure ices may be minimal or absent entirely.  

\subsubsection{Ganymede}
High-pressure ice phases V and VI, and possibly III, are likely present in Ganymede.   The planned JUICE mission \citep{grasset2013JUICE} will constrain the global density structure of Ganymede, its ice Ih thickness, and the conductance of its ocean.  Using this prior information, a seismic investigation could reveal the radial structures of the ices and search for any liquids within and between those layers, including the abundance of liquids at the water-rock interface. 

\subsubsection{Callisto} 
Poor constraints on Callisto's density structure mean that its ocean can be either Europa-like, less than 250~km deep and containing scarce or no high-pressure ices, or Titan-like, with seafloor pressures approaching 800~MPa and the presence of ices V and VI.  In either case, the fully stagnant surface suggests the ocean is nearly frozen and so is near-eutectic in composition.  This should lead to a high electrical conductivity. Buoyant ice III may be present, unless the ocean contains strong freezing suppressants such as ammonia or methanol. 

\subsection{Thermodynamic properties}
Examining the influence of composition on the geophysics and habitability of ocean worlds reveals which thermodynamic measurements are most needed. 

\textbf{Fluids}:  Evaluating the differences between high and low pH oceans  in larger satellites \citep[chloride vs sulfate;][]{zolotov2009chemical} requires high-pressure thermodynamic data for aqueous NaCl and mixtures with sulfates, including the associated melting point suppression of ices. Some of this information is  available at pressures relevant to watery exoplanets \citep{journaux2013influence,mantegazzi2013thermodynamic}, but crucial measurements in the 0.1-1~GPa range of pressures are lacking. 

The scope of the present electrical conductivity analyses is limited. A more rigorous treatment of conductivity should evaluate the limitations of available theory \citep{marshall1987reduced}. Laboratory electrical conductivity measurements  could test the extrapolations to high salinity assumed herein. 

\textbf{Solids}: To accurately interpret seismic measurements in terrestrial glaciers and in ocean worlds, sound speed measurements are needed in ices Ih, II, III, V, and VI at temperatures other than -35~$^o$C. This will also enable the development of thermodynamic properties comparable to those available for fluids, and more comprehensive predictions of the melting curves of ices in the presence of salts. Related work, already in progress, is revealing possible importance of trapped ions within ices \citep{fortes2010phase,journaux2013influence,journaux2017salt}. The associated geophysical properties of these materials would enable investigations (seismic, gravitational, etc...) that might be pursued by future exploration missions.

\section{Conclusions}\label{section:conclusions}
Calculations of the radial compositional structure of icy ocean worlds demonstrate the utility of geophysical measurements as test for habitability. 
We have focused on the relation between ice thickness and ocean composition, building on more limited prior work \citep{vance2014ganymede} to illustrate the unique 1D structures arising from the combination of melting-point suppression and density in single-component oceans containing aqueous MgSO$_4$, NaCl (as seawater), and NH$_3$.  We have established the ability of the PlanetProfile model to accommodate additional ocean compositions as suitable thermodynamic data become available. Although we have limited the present suite of models to those with thermally conducting rocky interiors too cold to allow melts to form, we have established the ability to evaluate the influence of the rocky interior by considering different mineral and elemental compositions, hydration states,  porosity, and heat content.  

Future work focusing on specific worlds can make use of new thermodynamic data for fluids and solids to evaluate the couplings between rock and ocean composition, and the presence of various ice phases. Similarly, the properties of the iron core can be varied in concert the rock composition to explore the parameter space of heat and metal content in the cores of Europa, Ganymede, and possibly Titan.

Model outputs are suitable as inputs to forward calculations of seismic propagation, gravitational moment of inertia, tidal Love numbers, and induced magnetic fields, and so can be used in the design,  planning, and data analysis for geophysical investigations.





\appendix
\section{Model Details}\label{appendixMD}
\subsection{Mass and Gravitational Moment of Inertia}
The bulk mass of the planet $M$ is the sum of nested spherical shells of radius $r$ (thickness $dr$) with constant density $\rho(r)$:
\begin{eqnarray}\label{eqn:M}
M=\int_{M}dm = 4\pi \int_{R_{sil}}^{R}\rho(r)r^{2}dr+\frac{4\pi}{3}\rho_{sil}R_{sil}^{3}\\ \nonumber
= M_{H_{2}O}+\frac{4\pi}{3}\rho_{sil}R_{sil}^{3}
\end{eqnarray}
where the $M_{H_{2}O}$ includes the mass of the low-pressure ice upper layer, the ocean, and the high-pressure ice layers. The moment of inertia is:
\begin{eqnarray}\label{eqn:C}
C = \int_{M}x^{2}dm= \iiint r^{4} \rho(r,\theta,\psi) sin^{3}(\theta)dr d\theta d\psi \\ \nonumber
 = \frac{8 \pi}{3}\int_{R_{sil}}^{R}\rho(r)r^{4}dr + \frac{8\pi}{15}\rho_{sil}R_{sil}^{5}\\ \nonumber
 =C_{H_{2}O}+\frac{8\pi}{15}\rho_{sil}R_{sil}^{5}
\end{eqnarray}
where $x$ is the distance of the element of mass ($dm$) from the spin axis.  Starting from the surface ($R_{sil}=R-dr$), the radius of the silicate interface is decreased by $dr$. The overlying mass ($M_{H_{2}O}$) and moment of inertia ($C_{H_{2}O}$) are recomputed with each step. The solution of the interface radius ($R_{sil}$) provides the silicate density ($\rho_{sil}$) from Eq. \ref{eqn:M}. The model uses this value to calculate the moment of inertia from Eq.~\ref{eqn:C} for specified values of the temperature ($T_b$) at the base of the ice I layer and specified values of fluid salinity. 
\subsection{Silicate Interior}
To predict the radii of the silicate/H$_2$O interface ($R_{sil}$) and iron core ($R_{iron}$), the model prescribes the densities of the silicates ($\rho_{sil}$) and the iron core ($\rho_{core}$). The moment of inertia is then computed from:
\begin{eqnarray}\label{eqn:C2}
M_{iron}=M-M_{H_{2}O}-\frac{4\pi}{3}\rho_{sil}\big(R_{sil}^{3}-R_{iron}^{3}\big) \\ 
 R_{iron}=\bigg(\frac{M-M_{H_{2}O}-\frac{4\pi}{3}\rho_{sil}R^{3}_{sil}}{\frac{4\pi}{3}(\rho_{iron}-\rho_{sil})}\bigg)^{1/3}\\
 C=C_{H_{2}O}+ \frac{8\pi}{15}\big(\rho_{sil}(R_{sil}^{5}-R_{iron}^{5})+\rho_{iron}R_{iron}^{5}\big)
\end{eqnarray}
If no iron core is present (as stipulated by the model), $R_{sil}$ is a function of $\rho_{sil}$ and $C$. In all cases, the model finds the range of values allowed by the bounded moment of inertia and chooses those corresponding to the nominal value of $C$. 

\subsubsection*{Thermal Conduction in Ices} 
Thermally conductive profiles assume temperature-dependent conductivity of the form $k=D/T$, with temperature represented as a function of depth $z$ as
\begin{equation}
T(z) = {T_{b}}^{\frac{z}{z_{b}}}\, {T_{o}}^{1 - \frac{z}{z_{b}}},
\label{eq:conduc}
\end{equation}
and corresponding heat flux at the ice I-water interface ($z_{b}$ and $T_{b}$),
\begin{equation}
q_{b} = D\frac{\ln{T_{b}/T_{s}}}{z_{b}}.
\label{eq:heat}
\end{equation}
Assuming $D=$632~W~m$^{-1}$ for ice~I  provides accuracy of better than 10\% in the range of temperatures from 100-270~K. Scalings other than $T^{-1}$ from \cite{andersson2005thermal} are refit to match the middle of the provided range of temperature with a loss of accuracy of less than 5\%. In the calculation of solid state convection, $k$ is computed from the fits recommended by \citet{andersson2005thermal}.

\subsubsection*{Thermal Conduction in Rock} 
The temperature profile in the rocky interior is computed by the same method as in \citet{cammarano2006longperiod}:
\begin{equation}
T(r) = T_0+\frac{\rho_{rock} H}{6 k}(a^2-r^2)+\bigg(\frac{\rho_{rock} H r^3_b}{3 k}- \frac{q_b r_b^2}{k}\bigg)\bigg(\frac{1}{a}-\frac{1}{r}\bigg)
\label{eq:rockthermal}
\end{equation}
Inputs to PlanetProfile are: $\rho_{rock}$,  the mean silicate density specified as an input: $H$, the rate of heat production per unit mass ($H=Q/M$, determined from input of Q in W~m$^{-2}$); and $k$, the thermal conductivity of the rock. The remaining variables are determined by the model:  $T_0$, the temperature at the top of the silicate layer; and $a$ and $r_b$, the radius at the top and bottom of the silicate layer, respectively.  The heat flow at the base of the rock layer, $q_b$, is set to zero. Thermal conductivity is the same as that specified by \citet{cammarano2006longperiod}: $k=4.0$~W~(m~K)$^{1}$. 

\subsection{Solid State Convection in Ices}

Solid state convection in ice Ih is from the parameterization \citet{deschamps2001thermal} based on numerical experiments and known material properties. The critical Rayleigh number 
\begin{equation}
Ra=\frac{\alpha \rho g \Delta T h^3}{\kappa \mu}
\label{eq:Ra}
\end{equation}
is determined at the characteristic temperature of the well-mixed convective interior of the ice:
\begin{equation}
T_{c} = B\bigg[ \sqrt{1+\frac{2}{B}(T_m - C)}-1\bigg], B=\frac{E}{2Rc_1}, C=c_2 \Delta T
\label{eq:Tc}
\end{equation}
in which $T_m$ is the characteristic melting temperature, specified as temperature at the ice-liquid interface, $E$ is the activation energy of the material (Table~\ref{table:materialProps}), $R$ is the ideal gas constant (8.314 J~mole$^{-1}$~K$^{-1}$), $c_1=1.43$, $c_2=-0.03$, and $\Delta T=T_m-T_{top}$. The Newtonian viscosity is
\begin{equation}
\mu = \mu_0 \exp\bigg[A\big(\frac{T_m}{T}-1\big)\bigg], A=\dfrac{E}{RT_m}
\label{eq:mu}
\end{equation}
with coefficient $\mu_0$ specified in Table~\ref{table:materialProps}.

The lower thermal boundary layer scales with Rayleigh number as $Ra_\delta = 0.28Ra^{0.21}$. The  boundary layer's thickness is 
\begin{equation}
\delta = \bigg[\frac{\mu_c \kappa}{\alpha \rho g(T_m-T_c)}Ra_\delta  \bigg]^{1/3} 
\label{eq:deltaTBL}
\end{equation}
and the corresponding heat flux is
\begin{equation}
Q=\frac{k (T_m-T_c)}{\delta}.
\label{eq:QTBL}
\end{equation}

The temperature of the overlying brittle lithosphere is scaled as $T_{lith} = T_{top} + 0.3\Delta T$
The thermal boundary layer thickness is
\begin{equation}
e_{TBL}=\frac{k_{ice}(T_{lith}-T_{top})}{Q}.
\label{eq:eTBL}
\end{equation}

\begin{table}[h]
\caption{Ice Material Parameters}
\centering
\begin{tabular}{|c|c|c|c|c|c|}
\hline
Crystalline   & $D^b$   	& $\nu^b$ & $E^c$& $K_S^d$ & $\mu^d$\\
Phase  		  & W~m$^{-1}$   &  Pa~s & kJ mol$^{-1}$& GPa & GPa\\ 
\hline
I & 	632	&5$\times 10^{13}$ & 60&	9.96	&3.6\\
II & 	418	&1$\times 10^{18}$ & 77&	13.89	&6.2\\
III & 	242	&5$\times 10^{12}$ & 127&	9.87	&4.6\\
V   & 	328	&5$\times 10^{14}$ & 136&	14.19	&6.1\\
VI  & 	183	&5$\times 10^{14}$ & 110&	18.14	&7.5\\
\hline
\multicolumn{6}{l}{$^a$adapted from \citet{andersson2005thermal} Table 1; $^b$\citet{durham2001rheology} Table 4; }\\
\multicolumn{6}{l}{$^c$\citet{deschamps2001thermal}; $^d$\citet{gagnon1990acoustic}}
\end{tabular}
\label{table:materialProps}
\end{table}


%
%
%


\begin{acknowledgments}
The PlanetProfile model and applications described in this work are available online (https://github.com/vancesteven/PlanetProfile), and support for its use may be obtained from the author via email. Work by JPL co-authors was partially supported by strategic research and technology funds from the Jet Propulsion Laboratory, Caltech, and by the Icy Worlds node of NASA's Astrobiology Institute (13-13NAI7\_2-0024). SDV acknowledges the support of NASA Outer Planets Research Grant NNH12ZDA001N "Solution Thermochemistry Relevant to Outer Planets and Satellites". RL acknowledges the support of NASA Outer Planets Research Grant NNX13AK97G “Physical Processes in Titan's Seas”. SCS was supported by grant SI1538/4-1 of Deutsche Forschungsgemeinschaft \textit{DFG}. No conflicts of interest are identified associated with this work. A part of the research was carried out at the Jet Propulsion Laboratory, California Institute of Technology, under a contract with the National Aeronautics and Space Administration.  Copyright 2017. All rights reserved. 
\end{acknowledgments}











\bibliography{BibTeX/svance}

\newpage

%
%
 
%






\end{document}